\documentclass{article}%
\usepackage{amsmath}
\usepackage{amsfonts}
\usepackage{amssymb}
\usepackage{graphicx}%
\providecommand{\U}[1]{\protect\rule{.1in}{.1in}}
\newtheorem{theorem}{Theorem}
\newtheorem{acknowledgement}[theorem]{Acknowledgement}

\begin{document}

\title{Causality and dispersion relations and the role of the S-matrix in the ongoing research\\{\small To the memory of Jaime Tiomno (1920-2011)}\\{\small To be published in "Foundation of Physics"}}
\author{Bert Schroer\\present address: CBPF, Rua Dr. Xavier Sigaud 150, \\22290-180 Rio de Janeiro, Brazil\\email schroer@cbpf.br\\permanent address: Institut f\"{u}r Theoretische Physik\\FU-Berlin, Arnimallee 14, 14195 Berlin, Germany}
\date{January 2011}
\maketitle

\begin{abstract}
The adaptation of the Kramers-Kronig dispersion relations to the causal
localization structure of QFT led to an important project in particle physics,
the only one with a successful closure. The same cannot be said about the
subsequent attempts to formulate particle physics as a pure S-matrix project.

The feasibility of a pure S-matrix approach are critically analyzed and their
serious shortcomings are highlighted. Whereas the conceptual/mathematical
demands of renormalized perturbation theory are modest and misunderstandings
could easily be corrected, the correct understanding about the origin of the
crossing property requires the use of the mathematical theory of modular
localization and its relation to the thermal KMS condition. These new
concepts, which combine localization, vacuum polarization and thermal
properties under the roof of modular theory, will be explained and their
potential use in a new constructive (nonperturbative) approach to QFT will be
indicated. The S-matrix still plays a predominant role but, different from
Heisenberg's and Mandelstam's proposals, the new project is not a pure
S-matrix approach. The S-matrix plays a new role as a "relative modular invariant"..

\end{abstract}

\section{Introduction to the various causality concepts along historical
lines}

Analytic properties of scattering amplitudes which arise as consequences of
causal propagation properties were first studied in the context of classical
optics in dielectric media and appeared first under the name
\textit{dispersion relations} in the late 20s in the work of Kramers and
Kronig\footnote{Here and in the following we refer references to the
bibliography in \cite{Moyses} wherever it is possible. This monography is a
competent and scholarly written account of the subject, though it does not
contain the QFT derivation which is based on the Jost-Lehmann-Dyson
representation, the latter can be found in \cite{I-Z}.}. The mathematical
basis on which this connection was derived amounted basically to an
application of Titchmarsh's theorem: a function (more generally a
distribution) $f(t)$ which is supported on a halfline, is the Fourier
transform of a function $a(\omega)$ which is the boundary value of a function
whose analyticity domain is the upper half-plane. With appropriate
restrictions on the increase at infinity, this analytic behavior can be recast
into the form of a dispersion relation (relating the absorptive to the
dispersive part) which is the best form for experimental checks of causality.
In the simplest case such a relation is of the form%
\[
a(\omega)=\frac{1}{\pi}\int_{-\infty}^{+\infty}\frac{Im~a(\omega^{\prime
})d\omega^{\prime}}{\omega-\omega^{\prime}},~~Im\omega>0
\]

Only after world war II this idea of relating spacetime causality with
analyticity in the form of dispersion relations found its way into quantum
theory (QT). Sch\"{u}tzer and Tiomno \cite{S-T} were among the first who
worked out conditions under which a dispersion relation can be derived for
scattering amplitudes of elementary processes in the setting of quantum
mechanic (QM); later there appeared other contributions with a different
adaptation of the notion of causality and slightly different restrictions on
the two-particle interaction potentials. These considerations can in principle
be extended to a more recent relativistic generalization of QM called
"\textit{Direct Particle Interaction}" (DPI) \cite{Coe}\cite{interface}, a
theory which is solely build on particles without using fields or algebras of
local observables. In such a setting, in which the Poincar\'{e} group is
unitarily represented and the S-matrix comes out to be Poincar\'{e}-invariant,
there is no implementation of micro-causality, similar to the work of
Sch\"{u}tzer and Tiomno one can only implement
macro-causality\footnote{Macro-causality is based on Born-localization and
refers primarily to wave functions and their large time propagation. It leads
to the concept of velocity of sound in QM and to the velocity v%
$<$%
c for relativistic mechanics (DPI). Micro-causality is an algebraic property
of local observables which entails spacelike Einstein causality and the
(timelike) \ causal completion property.} which includes the spacelike
cluster-factorization and Stueckelberg's \textit{causal rescattering}
requirements \cite{Stue}\cite{interface}. The additional difficulty in the DPI
case, which was solved in \cite{Coe}, is that the naive addition of pair
potentials in nonrelativistic QM would be in contradiction with the
multiparticle representation of the Poincar\'{e} group representation and
those macro-causality requirements, i.e. those coarse causality requirements
which can solely be formulated in terms particles.

The main problem in passing from classical optics to QM is that the latter has
no finite limiting velocity and admits no wave fronts. Wave packets dissipate
instantaneously in such a situation; a finite velocity (as the speed of sound)
in a quantum mechanical medium ( idealized e.g. as a lattice of oscillators)
arises only as an "effective" velocity of a disturbance; more precisely as an
asymptotically defined (large time) mean value in wave packet states. In that
case localization is related to the spectral theory of the \textit{position
operator} $\mathbf{x}_{op}$ and described in terms of a wave functions
$\psi(\mathbf{x})$ which is a square integrable function on its spectrum. The
position operator is frame-dependent (non-covariant) and enters particle
physics together with the Born-probability; although the probabilistic
"Born-localization" is an indispensable concept of particle physics, it is not
intrinsic to QM and has been (and still is) the point of interpretational and
philosophical contentions\footnote{The most prominent opponent against Born's
intoduction of the probabilistic aspect to QM was Einstein, even though the
resolution of the thermal aspect of the "Einstein-Jordan conundrum" which,
similar to the Unruh Gedankenexperiment brought thermodynamic probability into
zero temperature QFT \cite{hol} \ may have softened his resistence.}. There is
no way to talk about localized observables in QM; the obvious attempt to go to
the second quantized Schr\"{o}dinger formalism and define $\psi_{op}%
(\mathbf{x})$ and its local functions as pointlike local generating fields, or
to introduce region-affiliated observables by smearing, does not work. These
objects loose this property immediately since; unlike for wave function
propagation, there is no stable meaning of "effective" on the level of
localized observables; quantum mechanical algebras loose their localization at
a fixed time instantaneously.

In "direct particle interactions" (DPI) which implements interacting particles
in the setting of a multi-particle Poincare representation theory the
impossibility of reconciling the particle concept with covariance remains even
though with careful implementation of the cluster factorization one can at
least obtain a Poincar\'{e} covariant scattering matrix with the correct
macro-causality properties \cite{interface}. The problems of lack of
covariance of particle localization and dissipation of wave packets continue
in QFT; the probabilistic setting places the particle physics interpretation
of the velocity of light on the same conceptual level as that of the velocity
of sound, both are "effective" or asymptotic since there is no limiting
microscopic velocity.

In QFT one abandons attempts to implement properties of relativistic
localization with the help of particles; instead one adapts the idea of a
causal propagation of classical wave equations to the setting of QT and
postpones the relation with relativistic particles to a second pass. The
algebraic side of QFT comes with a very precise definition of causal locality
in terms of spacetime-indexed operator algebras $\mathcal{A(O})$; it consists
of two parts%

\begin{align}
\left[  A,B\right]   &  =0,~A\in\mathcal{A(O)},~B\in\mathcal{A(O}^{\prime
}\mathcal{)}\subseteq\mathcal{A(O})^{\prime},~Einstein~causality\label{c}\\
\mathcal{A(O})  &  =\mathcal{A(O}^{\prime\prime}%
),~causal~shadow~property,\text{ }\mathcal{O}^{\prime\prime}%
\ causal~completion~of~\mathcal{O}\nonumber
\end{align}
Here the first requirement is the algebraic formulation of the statistical
independence of spacelike separated observables; the upper dash on the
spacetime region $\mathcal{O}^{\prime}$ denotes the spacelike disjoint region
to $\mathcal{O}$, whereas on the algebra it stands for the commutant algebra.
The second line is the local version of the "time-slice" property \cite{H-S}
where the double causal disjoint $\mathcal{O}^{\prime\prime}$ is the causal
completion (causal shadow) of $\mathcal{O}$ i.e. the area of total dependence
of $\mathcal{O}~$which is the algebraic quantum counterpart of classical
hyperbolic propagation. The algebras $\left\{  \mathcal{A(O})\right\}
_{O\subset R^{4}}~$form a local not of algebras whose inductive limit is the
global algebra. The two causality requirements are not independent. If the
Einstein causality can be strengthened to Haag duality\footnote{The
exceptional cases where Haag duality for local observables breaks down are of
special interest. For the local algebras generated by the free Maxwell field
this happens for $\mathcal{O=}$toroidal spacetime region (the full QFT version
of the semiclassical Aharonov Bohm effect \cite{Sch1}).}
\[
\mathcal{A}(\mathcal{O}^{\prime})=\mathcal{A(O})^{\prime},~Haag~duality
\]
the causal shadow property would follow. In physically acceptable models
observable algebras violate Haag duality only for multiply connected regions;
a prominent illustration is given by the QFT Aharonov-Bohm effect
\cite{charge}.

Physically undesirable reasons why QFTs could violate the causal shadow
property may in particular occur in correspondences between models in
different spacetime dimensions. In that case the lower dimensional contains
all the degrees of freedom of the higher dimensional one which is way too much
and leads to a violation of the causal shadow property: $\mathcal{O}%
^{\prime\prime}$ contains more degrees of freedom than $\mathcal{O}$ so that
one would have the impression that additional degrees of freedom would have
from outside our living spacetime in a mysterious "poltergeist" manner. This
cannot happen in the setting of Lagrangian quantization; but since the latter
has remained outside mathematical control and only describes a small subset of
QFT anyhow (last section), it is indispensable to have a rigorous structural
definition outside a Lagrangian setting. The time slice principle \cite{H-S}
was introduced to exclude such unphysical aspects as those of certain
generalized free fields.

The causal shadow property does not lead directly to restriction on single
operators, but is expected to play a prominent role in securing a complete
particle interpretation of a QFT \cite{Ha-Sw}\cite{Haag}. Whereas properties
of particles reveal nothing about the more foundational properties of fields
and local observables, the latter should lead to the former. We will return to
this problem later on.

In its global form, namely for $\mathcal{O}=\mathbb{R}^{3}$ at fixed time or a
time-slice $0<t<\varepsilon,$ it has also been called \textit{primitive
causality}\footnote{The meaning of primitive causality in \cite{Moyses} is
slightly different.}. Whereas in this global form it exists also in QM, the
local (causal shadow) form is specific for QFT. Without the particle concepts,
field variables and the localized algebras which they generate would lead to a
mathematical "glass bead game" without much physical content; with the
exeption of the electromagnet field a quantum field is not directly
measurable, the intervention of particles is indispensable. It is of paramount
importance for the physical interpretation of QFT that the two notions of
localization, the particle-based \textit{Born-Newton-Wigner localization} of
wave functions and the algebraic notion of \textit{causal localization} of
local observables, which for finite times behave in an antagonistic way, come
harmoniously together in the timelike asymptotic region of scattering. This
implies in particular that the \textit{frame-dependent BNW- localization}
becomes \textit{asymptotically frame-independent}. Without this asymptotic
coexistence there would be no scattering probabilities and hence no particle
physics as we know it. Therefore instead of highlighting the misfit of the
particle localization concept with that of fields in QFT at finite spacetimes,
it is better to emphasize the asymptotic particle/field harmony which is after
all everything one needs (the half glass full against the half glass empty
view). It is remarkable that in d=1+3 interacting QFT not even a noncompact
spacetime localization region as large as a wedge permits the existence of
vacuum polarization free one-particle generators (PFG) with a reasonable
behavior under translations. Nevertheless in the last section we will present
a new approach which extends the intrinsic representation theoretic approach
to free particles and fields by Wigner to the realm of interacting QFT; the
new idea is: \textit{emulation} of wedge-localized Wick-products of the
incoming free field inside the wedge-localized algebra $\mathcal{B}(W)$
generated by the interacting fields. A special case, which is only available
in d=1+1, leads to the rich class of factorizing models\footnote{Even in those
few cases where they have a Lagrangian name, their existence and the
construction of their formfactors cannot be achieved by Lagrangian
quantizations.}.

The absence of the position operator among the observables of QFT also implies
that there is no conceptual basis for the derivation of Heisenberg's
uncertainty relation. The thermal KMS properties of the global vacuum state
after restricting it to the localized subalgebra $\mathcal{B(O})$ offer
however an algebraic substitute (see last section): the increase of
localization entropy/energy by passing from a "fuzzy" causal horizon (with
"attenuation size" $\varepsilon$ for the vacuum polarization cloud) to the
diverging limit $\varepsilon\rightarrow0$ (corresponding to infinite
uncertainty in QM). Interestingly the restriction of the vacuum to a state
resulting from the global vacuum also imports an intrinsic statistical
mechanics probability aspect into QFT which is independent from that by Born
which entered via scattering theory.

In addition to the above spacelike commutativity and causal completion
property, the spacetime-indexed local algebras fulfill a set of obvious
consistency properties which result from the action of Poincar\'{e}
transformations on the localization regions. This is automatically fulfilled
if these spacetime indexed operator algebras are generated by finite-component
Poincar\'{e} covariant fields $\Psi^{(A,\dot{B})}$ fields (where our notation
refers to the well-known dotted/undotted spinorial formalism).

As a result of the "ultraviolet crisis"\footnote{It was really a crisis
resulting from insufficient understandings about QFT.} of QFT, which started
in the 30s and lasted up to the beginnings of renormalization theory at the
end of the 40s, local QFT became discredited and the introduction of an
elementary length into QFT, or its total abandonment giving up localization in
favor of an ultraviolet-finite unitary S-matrix, were seriously contemplated.
Although QFT enjoyed a strong return, the longings for a S-matrix theory or a
nonlocal (often by inferring nonlocality via "non-commutativity") never
disappeared even though all these attempts during more than half a century
remained unsuccessful (section 3.1)

With doubts about the validity of the principle of causal locality, and in
spite of the observational success of perturbative QED in which these
principle is realized, there was a strong desire to directly check the
validity of the forward dispersion relation, which was done in 1967
\cite{test} for $\pi-N$ scattering at up to that time available energies. This
was the first (and the last) time a model-independent fundamental principle of
a relativistic QT was subjected to a direct experimental test in the presence
of strong interaction where perturbation theory\footnote{A perturbative
account of dispersion relations and momentum transfer analyticity was
presented in \cite{Tod}. At that time the divergence of the perturbative
series was only a suspicion, but meanwhile it is a fact.} was not applicable.
The importance of this successful test results from the rigorous and profound
mathematical-conceptual work which connected the causality principle to the
dispersion relation; for an unproven conjecture such a concerted effort could
hardly have been justified. Therefore the precise and detailed mathematical
work was a valuable investment in particle physics. Whereas in the early work
of quantum mechanical derivations of dispersion relations the latter were not
the consequence of a principle but rather properties of the chosen potentials,
the much more subtle derivation in QFT revealed that they are a direct
manifestation of its foundational causal locality principle. The link between
this principle and its analytic consequences was a spectral representation for
the particle matrix element of two fields, the so called Jost-Lehmann-Dyson
representation \cite{I-Z} which generalizes the simpler K\"{a}ll\'{e}n-Lehmann
representation for the two-point function to particle matrix elements of
commutators of two fields. It used to be well-known in the 60s and it would go
beyond the purpose of this paper to review important results of the past which
were lost in the maelstrom of time; for readers who have problems to get the
relevant information from the internet, I suggest to visit a library. As a
side remark, the JLD spectral representation has been almost exclusively used
for the derivation of analytic S-matrix properties; I only know of one quite
different use: the Ezawa-Swieca proof \cite{E-S} of the Goldstone's conjecture
based on the generalization of a property of a concrete Lagrangian model
stating that a conserved current yields a divergent global charge (spontaneous
symmetry breaking) only if a zero mass particle (the Goldstone boson) prevents
the large distance convergence\footnote{The other alternative, namely that the
global charge vanishes \cite{Sw} is related to the Schwinger-Higgs mechanism
of charge screening.}.

The formulation of the dispersion relations and their experimental
verification was a remarkable achievement in several respects. Besides the
aforementioned aspect of a confrontation of a principle directly with an
experiment without the intercession of a model, it is the only case of a
"mission accomplished" achievement in high energy physics. This is because
after the problem had been formulated within the setting of QFT and worked out
theoretically, it underwent a successful observational test; the physicists
who participated in this unique endeavour could afterwards turn their interest
to other problems with the assurance and feeling of satisfaction of having
contributed to the closure of one important problem.

It is important not to keep the dispersion relation project separate from
later attempts to base particle theory solely on the construction of an
S-matrix. The first such attempt, based on a postulated two variable spectral
representation of the elastic scattering amplitude was proposed by Mandelstam
in 1957 \cite{Mand}. I remember from my attending seminars at the University
of Hamburg that Lehmann Jost and K\"{a}ll\'{e}n were somewhat unhappy about
the increasing popularity of unproven representations. But the main purpose of
Mandelstam's representation was not to give an alternative support for
dispersion theory but rather to test the possibility of an S-matrix approach
to particle theory. In the present work we will not only explain how a
foundational understanding of causality leads to a new view of QFT (including
the origin of the important crossing property in particle physics), but we
will also convince the reader that a lot can be learned from an in-depth
understanding of the derailment of all pure S-matrix attempts. In doing this
we will learn that the S-matrix has a property which connects it with
"wedge-localization" and it is this recently discovered property which makes
it an indispensable constructive tool of QFT. In this sense, and only in this
sense (and not in the dual model/string theory attempts) Mandelstam's idea
about the central position of the S-matrix in particle theory survives.

Naturally a successful concluded project as the unravelling of the connection
between causality and dispersion relation invites to look out for extensions
into other directions. There were at least two good reasons for this. One was
that the successful perturbative approach in QED could not be expected to work
in strong interactions (at that time $\pi-N$ interactions). The other is more
profound on the theoretical side and relates to the at that time growing
suspicion that renormalized perturbative series in QFT may always diverge; a
suspicion which later on became a disturbing fact, since it meant that the
only known way to access Lagrangian interactions did not reveal anything about
the mathematical existence. This insight relativizes the success of QED
somewhat, because to realize that the only remaining possibility, namely
asymptotic convergence for infinitesimally small couplings has no useful
mathematical status, is a sobering experience. An experimental comparison with
perturbation theory is only fully successful if the theory has a
mathematical-conceptual existence status. This deficiency of the presently
only calculational access distinguishes QFT from any other area of theoretical
physics were sufficiently nontrivial soluble examples were available and
non-integrable models permit mathematically controlled approximations. The
tacit assumption underlying all quantum field theoretical research is of
course that this a temporary shortcoming of our capabilities and not a flaw in
our characterization of QFT.

The S-matrix boostrap project, which was vigorously proposed (notably by
Chew), was based on the conjecture that a unique S-matrix (primarily of strong
interactions), can be determined on the basis of three principles unitarity,
Poincar\'{e} invariance and the crossing property, where the last requirement
extends the analyticity of the dispersion relations. Viewed in retrospect, it
is not these three requirements which cause raised eyebrows, but rather the
idea that one can use them to "bootstrap" one's way into finding \textit{the}
S matrix of strong interactions. Such ideas about the existence of a unique
particle physics "el Dorado" which can be found by juxtaposing the right
concepts have arisen several times in particle physics; they are in fact
harbingers of the later millennium theory of everything (TOE). One of their
features is usually a highly nonlinear property as the unitarity requirement
in the bootstrap case. In such a situation "pedestrian" attempts to join
linear requirements with nonlinear ones lead in most cases to an explosive
nonlinear batch, to which no solution can be found by computational tinkering.

This negative result is sometimes not the end of the story since it may
nourish the hope (as in the similar case of the nonlinear Schwinger-Dyson
equations) that the principles only allow one solution which, as a result of
its uniqueness is very hard to construct. This is the basis of the belief in
the existence of a theory of everything (TOE), whereas in QFT the refinement
of physical principles only lead to a refined selection of models. The idea
that one can nail down a unique solution by physical principles implemented by
a bootstrap mechanism had some prominent supporters in the 60s besides its
protagonists Chew, Mandelstam and Stapp; also Dyson initially supported this
project. QFT avoids such direct encounters with nonlinear properties by
implementing unitarity via the asymptotic convergence of Hermitian fields in
the setting of scattering theory.

The invigorated QFT in the form of Yang-Mills gauge theories, and the lack of
concrete computational bootstrap problems caused a shift away from the
bootstrap project towards QFT; As we know nowadays, there is nothing wrong
with those bootstrap postulates, the error lies in the expectation that they
can be directly used in this raw form to implement calculations, they are
simply too vague (in particular the crossing property, see last section).

The few remaining adherents of an S-matrix approach who did not convert fully
to gauge theory, turned to more phenomenological motivated problems of strong
interactions as the use of the idea of Regge trajectories; this led them to
the class of \textit{dual resonance models }\cite{Mandelstam}, which
afterwards culminated in string theory. With discrepancies in the
observational scattering results involving high momentum transfer, the
phenomenological support evaporated. The whole setting was too sophisticated
for being supported by phenomenological applications and there arose the idea
to connect the orphaned mathematical formalism with more fundamental physics.
An argument based on formal manipulations of functional integrals catapulted
the phenomenological dual model formalism and its string theory extension into
the observational inaccessible setting of gravity at the Planck
scale\footnote{As cynics commented, this was the safest way to prevent any
further (after its failure in strong interactions) disagreement with
observations.}.

In retrospect it is a surprising that this found the unqualified support of
the dual model/string community since at that time the critical tradition in
particle theory had not yet completely disappeared and there was no common
accepted viewpoint of what to make out of the "dual model formalism" involving
properties of Gamma functions which was interpreted as a concretization of
Mandelstam's double spectral representation of the elastic scattering
amplitude. The two additional steps which show that the crossing property of
particle theory was not the same as in Veneziano's dual model are one
implemented in the dual model are more recent. The first one was the
recognition that the formalism of the dual model is based on a crossing
property of the Mellin transform of conformal 4-point functions \cite{Mack}%
\cite{foun}. This will be explained in section 3.1. The true crossing property
involving particle states on the other hand follows from the thermal KMS
properties of the wedge localized interacting algebra together with that of
the corresponding free algebra which shares the same modular group. The
derivation is based on deep properties of modular localization and will be
explained in section 3.2 The crossing property for the S-matrix is a special
case of the formfactor crossing since the transition amplitudes are
formfactors of the identity operator (section 3.2).

Since causal localization is not only the most important but also the subtlest
of all properties of a relativistic QT it is not surprising that errors were
committed in the various attempts of its implementation. What is however
difficult to understand why it was not possible to repair them within more
than 4 decades. Certain scientific developments cannot be understood within
the realm of science since they are of a sociological nature. It is evident
that an area of research as particle theory, which is by its very nature quite
speculative on its unsecured frontier, needs a strong critical corrective for
keeping it on track. At the times of competing schools (Landau, Bogoliubov,
Lehmann, Jost, Haag) and independent strong individuals as Pauli, the
"Streitkultur" in the old world was keeping the particle theory caravan on
track and was able to hold course even against some at the time popular
fashions which originated in the new word. All this changed when the competing
schools were replaced by globalized communities and instead of a Streitkultur,
practiced by renowned particle physicists, there was now the unchallenged
suggestive power of the reputable community leaders. To make things worse, the
newcomers who enter the monocultures practised within a globalized community
learn the computational tools in order to become productive within the
community, but have no chance to get to that level of critical knowledge which
is required to get out of the derailment. This may very well continue into the
post LHC situation. Although this is an interesting and important issue to be
pursued, our criticism in this paper will be kept on the
conceptual-mathematical track. The lack of observational success of string
theory which has been the critical focus in several articles, plays no role
here. Rather the mere conceptual existence of a trans-QFT setting in
relativistic QT which goes beyond QFT in the sense of incorporating its
localization concepts as a limiting case, would by itself be a remarkable
discovery, independent of its observational status. QFT is the first
completely intrinsic and holistic theory \cite{hol} (see later) to which for
more than 70 years no alternative has been found; every attempt (including
S-matrix theories) to find an consistent alternative has failed.

These critical remarks need to be somewhat qualified by emphasizing the recent
constructive progress on QFT\ for theories with an asymptotically complete
particle interpretation that attributes a crucial role to the S-matrix.
Besides its well-known role in scattering theory it carries some information
about wedge-localization ($S_{scat}$ is a relative modular invariant). In fact
it is precisely this role which allows to formulate the new construction
"doctrine" of starting QFT constructions with wedge-localized generators. This
is the reverse of the standard doctrine in QFT which is based on local
interactions between pointlike free fields and passes to more global
constructions including the scattering amplitudes. The advantage of pointlike
objects as fields is that they are in the range of the quantization
parallelism of classical field theory. But there is a high prize for this to
be paid since the analogy to classical fields ignores the QFT phenomenon of
localization-caused vacuum polarization. These vacuum polarizations
intuitively speaking tend to enforce "Murphy's law": channels which can couple
(according to superselection rule) actually will couple. This makes QFT (in
contrast to QM) a fundamental theory, but it also renders it inaccessible to
functional analytic (single operator) methods. The important new lesson is
that generators of wedge algebras have much simpler vacuum polarization
properties than quantized pointlike fields; in fact wedge generators present
the best compromise in the somewhat antagonistic quantum particle-field relation.

This philosophy was very successful for a class of models which before were
treated with the nonperturbative bootstrap formfactor prescription \cite{Ka}.
Similar to the dispersion relation project, it accomplished all its self-posed
aims. But as a result of its more modest ambition and guarded presentation, it
remained little known to the majority. For this reason it may be helpful to
use the remainder of the introduction to present same details about their
fascinating history. This models are the only integrable QFTs\footnote{They
may be seen as the QFT analogs of the Kepler- or quantum mechanical hydrogen-
problem, but whereas integrable systems in the classical or quantum mechanical
setting exist in any spacetime dimension, integrability in QFT does not extend
beyond d=1+1 \cite{integrable}.} and as a result of the structure of their
S-matrix which factorizes into products of an elastic two-particle scattering
function, they are also called \textit{factorizing models.}

They came into being through the observation that certain quasiclassical
aspects, first noted by Dashen, Hasslacher and Neveu on some 2-dimensional
QFTs (the most prominent was the Sine-Gordon equation) suggested that the
integrability was related to the bootstrap properties of their two-particle
elastic S-matrices in particular the nuclear democracy principle of its
one-particle structure \cite{STW}. The existence of an infinite family
contradicted of course the TOE uniqueness which was part of the bootstrap
doctrine. Even worse for the bootstrap ideologues, each of the purely elastic
two-dimensional S-matrices was associated to the system of formfactors of a
unique (in an appropriate sense) QFT \cite{Ka}.

The bootstrap-formfactor construction was enriched on the algebraic side by
the two brothers Zamolodchikov \cite{Zam}. I found this new tool quite
intriguing and since in my understanding the characterizing concept of QFT is
locality, I looked for a spacetime interpretation of these operators and
realized that they are better than simply nonlocal, their Fourier transform is
wedge-local \cite{AOP} and this property makes them much more than an mnemonic
device; in the form of vacuum-polarization-free generators \cite{BBS} they
become a new constructive tool in the setting of algebraic QFT. With their
help it was possible to show the true (nonperturbative)
conceptual-mathematical existence of interacting factorizing models
\cite{Lech} \cite{Lech2}. After 80 years absence of mathematically control
here was a class of two-dimensional models with noncanonical short distance
behavior for which all doubts about their existence could be removed and many
of the objects one is interested in were computed. The computational
difficulties turn out to be opposite to those of perturbation theory, namely
the more one moves off mass-shell the more extensive the calculations become:
the S-matrix and the formfactors require less computational work, whereas the
correlation functions remain prohibitively complicated. Fortunately the
existence proof does not require such explicit calculations and neither does
the extraction of physical properties as formfactors require the knowledge of
correlation functions of fields. For the first time the Lagrangian
quantization and functional integral approach in which a less fundamental
(classical) theory is supposed to give the tune according to which a
fundamental QFT has to dance has been turned around; for most of the models a
Lagrangian is not known and there are qualified doubts that a Lagrangian
"baptism" exists\footnote{The cardinality of scattering functions obeying the
bootstrap principles is bigger than that of renormalizable Lagrangian
couplings of free fields.}. This is certainly a respectable success after many
decades of stagnation, even if the limitation to d=1+1 still reminds us of the
enormous work ahead which will be necessary in order to be able address these
problems in realistic d=1+3 models.

These constructions of low dimensional QFTs have led to renewed interest in
the origin of the crossing property which plays a crucial role in S-matrix and
the more general formfactor properties. Some of the ideas, especially those
about the conceptual origin of the crossing property of formfactors, combined
with the progress in local quantum physics (LQP), have led to a surprising
connection between crossing and the thermal aspects of modular localization.
This gave rise to a new setting of QFT in which the S-matrix plays a new
constructive role within QFT. It turned out that the analytic crossing
identity is related in a deep way to the KMS identity which expresses the
thermal aspect of the vacuum restricted to the interacting wedge algebra.
Since the KMS property refers to only one algebra one has to "emulate" the
wedge localized free fields inside the interacting algebra \cite{foun}%
\cite{J+S}. The aforementioned setting of two-dimensional factorizing models
re-appears in this new setting as a special (in fact the only) case in which
the emulated operators have simple properties under all translations whereas
in the generic case the covariance of the emulated objects does not extend
beyond the transformations which map the wedge into itself. In explaining the
origin of the particle physics crossing property, the emulation process also
unravels an old problem in scattering theory about what happens if the wave
packets of incoming particle overlap \cite{Bu}\cite{Bu-Su}. In this case the
threshold singularities do not only limit the validity of the Haag-Ruelle and
LSZ scattering theory, but they also lead to radical modification of crossing properties.

This new setting could be considered as a heir to the old S-matrix project.
The crossing property plays a crucial role in both, but now not as a God-given
rule abstracted from Feynman diagrams, but rather as a fundamental consequence
of the thermal properties of modular localization. In this way a property from
the center of particle physics as crossing is conceptually united with the
Unruh effect\footnote{Also the thermal Hawking effect is localization-caused,
in this case the localization boundary is defined in terms of the curved
spacetime metric.} and with the recent complete understanding of the
"Einstein-Jordan conundrum" \cite{hol} at the cradle of QFT. This also adds a
philosophical touch to these new ways of looking at particle physics.

The new use of the S-matrix however destroy most of the old dreams about the
existence of a unique S-matrix theory which can be "bootstrapped" from some
postulates (the precursor of the later idea of a TOE). However it does render
the S-matrix and particle states important concepts to be used right at the
beginning of constructions of QFT, which goes somewhat (but not completely)
against Heisenberg's comment "the S-matrix is the roof of the theory and not
its foundation" \cite{EPJH} with which he distanced himself from his 1946
S-matrix proposal\footnote{Only in factorizing theories their purely elastic
S-matrices (only vacuum polarization no on-shell particle creation through
scattering) can be computed through the bootstrap project. In higher
dimensions there are no theories with only elastic scattering (\'{A}ks
theorem), real particle creation and vacuum polarization go together and
prevent a bootstrap construction.}. This is very different from the present
way of dealing with QFTs (e.g. perturbation theory) in which the vacuum
correlations of point-like fields of the canonical Lagrangian- or functional-
quantization are the main computational tools and only afterwards their large
time asymptotic limits or their momentum space mass shell projection are
constructed. It should be seen as the heir of the famous "Causality-Dispersion
relation project" of the 60s which, although studying on-shell objects, never
considered itself a pure S-matrix project.

\section{Macro- and Micro-causality}

The first S-matrix proposal for the construction of relativistic QTs in
1943/46 by Heisenberg \cite{Hei} was motivated by the desire to overcome the
reputed ultraviolet problem as well as to avoid the conceptional difficulties
of introducing a short distance behavior improving elementary length into QFT;
in a pure global S-matrix setting one would have gotten rid of the two
problems. Heisenberg's idea was that one may find sufficiently many properties
of $S$ directly i.e. without having to "interpolate" the incoming and the
outgoing particles in a scattering process by interacting (off-mass-shell)
pointlike local fields. There was no problem to account for the obvious
properties as unitarity, Poincar\'{e} invariance and the cluster factorization
for large spacelike separation%
\begin{align}
&  S=e^{i\eta},\text{ }e.g.~\eta=\int\eta(x_{1},..x_{4}):A_{in}(x_{1}%
)..A_{in}(x_{4}):d^{4}x_{1}..d^{4}x_{4},~\eta(...)\text{ }con.\\
&  \curvearrowright\underset{a\rightarrow\infty}{~lim}S(g_{1}..g_{k+1}%
^{a}..g_{n}^{a};f_{1}..f_{l+1}^{a}..f_{m}^{a})=S(g_{1}..g_{k};f_{1}%
..f_{l})S(g_{k+1}..g_{n};f_{l+1}..f_{m})\nonumber
\end{align}
Here a function $\eta(x_{1}...x_{n})$ is called connected ($con.$) if it
vanishes in the limit of large relative spacelike separation of $x^{\prime}s$;
$~f,g$ are wave functions and the upper $a$ denotes translation by $a.~$The
cluster property for $S~$(second line) is an immediate consequence of the
connectedness of $\eta.~$Unitarity is trivially satisfied by writing the
S-matrix in form of a Hermitian phase operator $\eta$ and the operational
Poincar\'{e} invariance follows from that of the coefficient functions $\eta$
(in general an infinite series), and the cluster property is a consequence of
the connectedness of the $\eta^{\prime}s.$ But as Stueckelberg pointed out
some years later \cite{Stue}, such an Ansatz lacks the macro-causality
property which he identified with an S-matrix property called "causal
rescattering" \cite{interface}. Unlike the three previous properties this
property has not and probably cannot be implemented " by hand"; it is
automatically fulfilled in models of QFT which have a complete particle
interpretation and it also can be implemented in the quantum mechanical direct
DPI models \cite{interface} mentioned in the sequel. In its simplest version
it states that the 3$\times$3 S-matrix should contain a particle pole
contribution which corresponds to a two-step process: first two of the
particles interact and then one of the outgoing particles interacts with the
(up this point) noninteracting third incoming particle.

That the second process happens an infinite time \textit{afterwards} means
that there is (in the S-matrix idealization) a pole term corresponding to the
timelike connection between the two 2-particle processes which has the same
$i\varepsilon~$prescription as the Feynman propagator; the only distinction is
that in the present case the latter has only asymptotic validity (in momentum
space near the pole). This "causal re-scattering" is an additional requirement
on S introduced by Stueckelberg which apparently cannot be implement "by hand"
while maintaining unitarity\footnote{It is automatically fulfilled (the causal
one-particle structure) in microcausal QFT and can be implemented in the
appropriately formulated (see below) DPI setting of relativistic QM.}. This
shows that a pure S-matrix theory without using a field-like mediator between
incoming and outgoing scattering states is not a realistic goal, a conclusion
that also Heisenberg reached some years later \cite{EPJH}. But the first
S-matrix attempt was not totally in vain, because by Stueckelberg's suggestive
ad hoc simplification of using the Feynman propagator also outside the
timelike asymptotic region, and assuming that the interaction region can be
shrunk to a point, he independently obtained the Feynman rules through
overidealizing macro-causality. So if Feynman would not have found an
operational setting for their derivation, we would not have been left
completely empty-handed since there would have been a perturbative suggestion
by Stueckelberg; however to prove perturbative on-shell unitarity without an
operational formalism is not an enterprise whose successful accomplishment is
guarantied. According to an article by Wanders (in \cite{Stue})
St\"{u}ckelberg actually found an iterative causal unitarization, starting
with a Hermitian Wick-product of fields and invoking micro-causality in every
iterative order to restrict the freedom to the structure of counterterms.
"Unitarization" by itself is not a well-defined procedure. At the moment one
invokes microcausality one has left the realm of a pure S-matrix theory. The
perturbative S-matrices one obtains this way are therefore identical to those
from a QFT, in fact the Epstein-Glaser perturbation theory is the
mathematically polished form of the Stueckelberg causal unitarization.

Macro-causal structures in scattering amplitudes (without their micro-causal
counterparts) are automatically fulfilled in theories with a spacetime
dynamics e.g. a Hamiltonian or an equation of motion. The "primitive
causality" in Nussenzveig's presentation of nonrelativistic scattering
problems \cite{Moyses} is based on the same physical idea, except that (unlike
causal re-scattering) one cannot remain within a pure S-matrix setting; the
definitions of Sch\"{u}tzer and Tiomno \cite{S-T} use the interaction dynamics
for all times. Such macro-causality concepts are quite efficient if one wants
to show that as hoc (not covered by principles) proposals of "modifications by
hand", as e.g. the introduction of the Lee-Wick complex poles into the Feynman
rules, lead to time precursors and in this way violate primitive causality
\cite{Swieca}.

These causality properties can be \textit{formulated} in terms of particles,
but can they also be computational \textit{implemented} in a pure particle
setting i.e. in a dynamics which is formulated \textit{only in terms of
particles}? There exists a little known quantum mechanical relativistic
multiparticle scheme\footnote{Although Dirac introduced important concepts
based on his project of a relativistic particle theory his implementation of a
particle-hole theory led to inconsistencies in perturbative orders in which
vacuum polarization entered.} which leads to interacting multi-particle
representations of the Poincar\'{e} group and fulfills all the macro-causality
properties which one can formulate in terms of interactions between particles
only: the setting of \textit{direct particle interaction} (DPI) \cite{Coe}.
\ Assuming for simplicity identical scalar Bosons, invariant energy operator
in the center of mass (c.m.) of two identical particles is $2\sqrt{p^{2}%
+m^{2}}~$and the interaction is introduced by adding an interaction term $v$%

\begin{equation}
M=2\sqrt{\vec{p}^{2}+m^{2}}+v,~~H=\sqrt{\vec{P}^{2}+M^{2}},P=p_{1}%
+p_{2},~p=\frac{(p_{1}-p_{2})_{c.m.}}{2}%
\end{equation}
where the invariant potential $v$ depends on the relative c. m. variables
$p,q$ in an invariant manner i.e. such that $M$ commutes with the Poincar\'{e}
generators of the 2-particle system which is a tensor product of two
one-particle systems.

One may follow Bakamijan and Thomas (BT) \cite{BT} and choose the Poincar\'{e}
generators in a way so that the interaction only appears explicitly in the
Hamiltonian. Denoting the interaction-free generators by a subscript $0,$ one
arrives at the following system of two-particle generators%
\begin{align}
\vec{K}  &  =\frac{1}{2}(\vec{X}_{0}H+H\vec{X}_{0})-\vec{J}\times\vec{P}%
_{0}(M+H)^{-1}\\
\vec{J}  &  =\vec{J}_{0}-\vec{X}_{0}\times\vec{P}_{0},~\nonumber
\end{align}
where the two particle operators $\vec{X}_{0},\vec{P}_{0},\vec{J}_{0}$ with
the subscript zero are just the sum of the corresponding one-particle
operators. The interaction $v$ may be taken as a \textit{local} function in
the relative coordinate which is conjugate to the relative momentum $p$ in the
c. m. system; but since the scheme anyhow does not lead to local differential
equations, there is not much to be gained from such a choice. The Wigner
canonical spin $\vec{J}_{0}$ commutes with $\vec{P}=\vec{P}_{0}$ and $\vec
{X}=\vec{X}_{.0}$ and is related to the Pauli-Lubanski vector $W_{\mu
}=\varepsilon_{\mu\nu\kappa\lambda}P^{\nu}M^{\kappa\lambda}$ .

As in the nonrelativistic setting, short ranged interactions $v$ lead to
M\o ller operators and S-matrices via a converging sequence of unitaries
formed from the free and interacting Hamiltonian \cite{Coe}%
\begin{align}
\Omega_{\pm}(H,H_{0})  &  =\lim_{t\rightarrow\pm\infty}e^{iHt}e^{-H_{0}t}\\
\Omega_{\pm}(M,M_{0})  &  =\Omega_{\pm}(H,H_{0})\label{sec}\\
S  &  =\Omega_{+}^{\ast}\Omega_{-}\nonumber
\end{align}
The identity in the second line is the consequence of a theorem which say that
the limit is not affected if instead of $M$ one takes take a positive function
of $M$ (\ref{sec}) as $H(M),$ as long as $H_{0}$ is the same function of
$M_{0}.$ This insures the asymptotic \textit{frame-independence }%
(P-invariance)\textit{ of asymptotic objects as the M\o ller operators and the
S-matrix, }but not that of semi asymptotic operators as formfactors of local
operators between \textit{ket} in and \textit{bra} out particle states. Apart
from this \textit{identity for operators and their positive functions}
(\ref{sec}), which seems to plays no role in the nonrelativistic scattering,
the rest behaves just as in nonrelativistic scattering theory. As in standard
QM, the 2-particle cluster property is the statement that $\Omega_{\pm}%
^{(2)}\rightarrow\mathbf{1,}$ $S^{(2)}\rightarrow\mathbf{1,}$ i.e. the
scattering formalism is identical. In particular the two particle cluster
property, which says that for short range interactions the S-matrix approaches
the identity holds also for the relativistic case if one separates the center
of the wave packets of the two incoming particles. Having a representation
theory of the two-particle Poincar\'{e} group does not imply that there are
covariant local observables, but together with the short range requirement
they secure at least the existence of a unitary Poincar\'{e} invariant two
particle S-matrix which obeys all macro-causality properties in terms of particles.

There is no problem in finding restrictions on the interaction $v$ which
correspond to those which e.g. Sch\"{u}tzer and Tiomno \cite{S-T} used in the
nonrelativistic setting. It is however nontrivial to generalize this setting
to \textit{multiparticle} interactions since the representation theory of the
Poincar\'{e} group prohibits a trivial implementation of cluster factorization
by adding up two-particle interactions as in the nonrelativistic case. The
Coester-Polyzou formulation of DPI shows that this is nevertheless possible
\cite{Coe}. The proof is inductive and passes the clustering of the n-particle
S-matrix to that of the n-particle Poincar\'{e} group representation which
than in turn leads to the clustering of the (n+1)-particle S-matrix etc. There
always exist unitaries which transform BT systems into cluster-separable
systems \textit{without affecting the S-matrix}. Such transformations, which
are unfortunately not unique, are called \textit{scattering equivalences.
}They were first introduced into QM by Sokolov \cite{So} and their intuitive
content is related to a certain insensitivity of the scattering operator under
quasilocal changes of the quantum mechanical description at finite times. This
is reminiscent of the insensitivity of the S-matrix against local changes in
the interpolating field-coordinatizations in QFT\footnote{In field theoretic
terminology this means changing the pointlike field by passing to another
(composite) field in the same equivalence class (Borchers class), or in the
setting of AQFT by picking another generator from the same local operator
algebra.} in QFT by e.g. using composites instead of the Lagrangian field.
From the construction it is clear that this relativistic DPI has no
fundamental significance. Its theoretical value consists in providing
counterexamples to incorrect conjectures as e.g. the claim that Poincar\'{e}
invariance of the S-matrix and cluster factorization requires QFT. Its
existence sharpens the recognition of the importance of the causal
localization and the depth in the particle-field dichotomy of QFT.

\section{Analyticity as a starting point for a theory?}

The two-fold limitation of perturbation theory resulting on the one hand the
divergence of its perturbative series and the ensuing doubts about the status
of existence of interacting QFT, and on the other hand its limitation to weak
couplings and the resulting problems for the description of nuclear
interaction (in those days the $\pi-N$ interactions) led to a return of the
S-matrix idea. In many aspects the new bootstrap ideas went beyond the
Heisenberg program and its criticism by Stueckelberg, but some of the old
conclusions were forgotten. The nontrivial macro-causality properties as the
spacelike clustering and the timelike causal rescattering property do not
anymore occur in the bootstrap list. Unimportant or forgotten in the maelstrom
of time? It is characteristic of the three historical S-matrix projects
(Heisenberg/Stueckelberg, bootstrap, dual model/superstring) that each
subsequent one added a new idea, but also ignored some of the older messages.
As will be seen in more details, the new post renormalization \textit{crossing
property} which was not present in the Heisenberg/Stueckelberg setting was
simply abstracted from analytic properties of Feynman graphs, but the lack of
understanding its physical root-causes carried the seeds of a conceptual
derailment whose far-reaching consequences strongly influenced the present situation.

Crossing is most clearly formulated in terms of formfactors, the crossing for
scattering amplitudes is a consequence of the formfactor crossing and the LSZ
reduction formula. Our proof in the formfactor setting (see last two sections)
is very different in scope and physical concepts from the proof for the
elastic scattering amplitude derived in the setting of axiomatic QFT using
techniques of functions of several complex variables \cite{BEG}\cite{Martin}.
It contains an element of surprise, since the so-called \textit{crossing
identity} turns out to be a somewhat camouflaged KMS identity which results
from the restriction of the global vacuum to the wedge localized algebra. That
the restriction of the global vacuum to the wedge algebra leads to an impure
thermal state which is KMS with respect to the wedge preserving Lorentz boost
is of course known from the \textit{Unruh and Hawking \ thermal manifestations
of localization} of quantum matter behind causal- or black hole event-
horizons\footnote{Contrary to popular opinion it is not the curvature but
rather the localization which generates the thermal aspect. The event horizon
attributes to the localization in front of the Schwarzschild horizon a
physical reality whereas the causal horizon of a Rindler wedge has a more
fleeting existence.}. But the manifestation of the thermal KMS identity
(implying that the $\mathcal{A(O)}$ -restricted vacuum is impure in a very
strong sense\footnote{For sharp localization it does not correspond to a
density matrix; only if one passes to a somewhat fuzzy surface the algebra
becomes a standard $B(H)$ algebra and the KMS state passes to a Gibbs density
matrix state. We believe that there is a relation to 't Hooft's "brick-wall"
construction \cite{brick}; further work on this poinr is required.}) in the
form of the particle crossing identity is somewhat unexpected (expressed in
the German subtitle of section 3.2). Further illustrations of this point can
be found in \cite{hol} \cite{cross}.

Analyticity as a postulate standing next to the other physically motivated
requirements was an important part of the Chew-Mandelstam S-matrix program
(the postulate of "maximal analyticity"). One does not have to share however
Chew's and Stapp's opinion that is by itself a physical principle. Rather it
is a consequence of the spectrum property and the causal localization
principle of QFT, but sometimes the path from the principle to the analytic
properties is subtle and demanding. To obtain an impression of the subtleties
one only has to look at the details of the derivation of the high energy
physics dispersion relation via the JLD representation from the causal
locality principles. The title of this section is a characterization of the
beliefs at the time of the Chew-Mandelstam S-matrix setting and the answer to
the question mark is: yes it is a powerful tool but only if one is able to
trace it back to its conceptual origin.

This section will be subdivided into three subsections. The first, entitled a
cul-de-sac, critically reviews the post bootstrap S-matrix project which
started with a concrete conjecture by Mandelstam about the existence of a
double spectral representation for the two variable crossing symmetric
scattering amplitude. The dual model resulted from a implementation of a
particular crossing property which is however not that of particle physics.
There are precisely two notions of crossing, that of crossing of conformal
4-point functions by applying the convergent global operator expansion
(established in conformal QFT) to the 3 different pairings of fields. In that
case the Mellin transform of this expansion produces a sum over infinitely
many poles whose position is defined in terms of the anomalous dimension of
the composite fields in the expansion. The correct particle crossing, which
will be presented in the third subsection 3.3, has nothing to do with this
conformal field crossing.

The bootstrap approach was given up, as a result of its incapacity of
producing reasonable calculations (it hardly led to any PhD thesis). Its heir,
the dual model, was more concrete and, as a result gave rise to many
calculations, especially after it became more phenomenological oriented with
the attempted incorporation of Regge poles and their trajectories. There
remains of course the interesting question: what would have happened if the
discovery of the dual model would have occurred in the clear daylight of the
Mellin transformed conformal field crossing instead to tinkering with
properties of gamma functions; would the dual model still have been considered
as a part of the scattering aspects of particle theory?

The use of the dual model for strong interactions was not only given up as a
consequence of disagreements with improved scattering data. Its highly
sophisticated mathematics was also conceptually out of tune with its
phenomenological aims. In this situation the idea that it could be useful in
Planck scale physics as a theory of quantum gravity gained ground.

The ultimate step to string theory resulted from an attempt to lend conceptual
physical importance to a mathematically tempting formalism by simply
forgetting the failed observational connection and postulating the yet unknown
gravitational physics at the Planck scale as its new range of application. Our
interest in this paper is limited to its problematic relation with the
localization concepts of QT, in particular whether it really has the
localization property which string theorist ascribed to it \cite{foun}.

The subsection 3.2 explains why the vacuum state restricted to a spacetime
localization region $\mathcal{O}$ turns into an impure thermal KMS state with
a vacuum-polarization cloud hovering in the vicinity of the causal/event
horizon\footnote{The localization-induced vacuum polarization may be seen as
the metaphor-free aspect of the "broiling vacuum polarization soup" of the
books on QFT \cite{Sum}.} \cite{BMS} \cite{interface}; the most interesting
case for testing foundational properties of QFT is $\mathcal{O=~}W$ (the wedge
region). It also provides additional insight into what physical
string-localization really means and why the objects of string theory are not
string- but rather point-localized. It also places big marks of doubts on the
string S-matrix proposal resulting from pure prescriptive manipulation of
functions which by decree are promoted to be scattering amplitudes. Without
being able to formulate these recipes in terms of states and operators they
are totally unconvincing. Rules which ignore the lesson learned from
Stueckelberg's iterative S-matrix construction by an operational unitarization
(see previous section) and which simply replace the worldline interpretation
for momentum space particle propagators by tubes (worldsheets) without backing
up such graphical analogies by Hilbert space operators fall back behind
Stueckelberg's work.

Even if this (after more than 4 decades) would still work by some overlooked
magic, there is still the conceptual problem of attributing a meaning to a
"stringy" S-matrix; an S-matrix is a \textit{global} object which a priory
does not contain any information about spacetime localization. In fact for the
only known case of \textit{genuine string-localization} as the best possible
localization, namely electrically charged matter fields \cite{Sch1}%
\cite{Sch2}, the consequence of the weaker than pointlike localization is the
(since Bloch and Nordsiek well-known) phenomenon of infrared divergences in
scattering theory which leads to the abandonment of the S-matrix in favour of
photon-inclusive cross sections (for which unfortunately no elegant LSZ like
representation in term of spacetime correlations has yet been found).

The second subsection of section 4 presents a new constructive setting in
which an algebraic version of crossing and \textit{analytic exchange}
(explained there) is the starting point of a new constructive approach which
in d=1+1 results in an existence proofs and the explicit construction of
formfactors for the class of factorizing models. In a certain sense this
success vindicates at least some aspects of the aspirations of the old dream
of the bootstrap community and in particular of Mandelstam \cite{Mand}
concerning the importance of analytic properties of on-shell objects, even
though its implementation requires quite different concepts as well as a
return to QFT. Nevertheless the use of the S-matrix as a basic computational
tool (and not just the roof of local particle physics) is shared with the
S-matrix attempts of the 50s and 60s which now, together with formfactors of
fields and a new much more subtle role of the crossing property, forms the
backbone of the new approach.

\subsection{Important lessons from a cul-de-sac}

The first two attempts to avoid fields in favour of a pure S-matrix approach
failed basically for the same reasons. The nonlinear unitarity of the S-matrix
together with other physically motivated linear requirements results in a
rather unwieldy computational batch. In the eyes of some created the
impression that if such a system of requirements admits any solution at all,
then it should be rather unique. In this way an early version of a theory of
everything (still without gravity) was born. But the disparity between high
dreams and the difficulty to translate them into credible computations led to
an early end of the projects. Young newcomers who were looking for doable and
credible computations were better served by the new nonabelian gauge theories
for weak and strong interactions.

The dream of uniqueness of the bootstrap ended at the beginning of the 70s
with the little noticed construction of an infinite family of elastic
scattering functions in d=1+1 \cite{Ka} which fulfill the bootstrap
requirements. But in contrast to the missionary zeal (if not to say cleansing
rage) against QFT by the adherents of the bootstrap project, this modest
observation about \textit{factorizing models} showed that each S-matrix came
precisely with one QFT whose scattering it describes. This relation
strengthened the idea of a unique association of a QFT to an S-matrix (i.e.
the uniqueness of the inverse problem of scattering theory in QFT)\footnote{An
S-matrix does however not distinguish a particular field, rather it associates
to a local equivalence class (Borchers class) or more compactly to a unique
net of local operator algebras of which those fields are different
generators.}. Most of the factorizing models, whose construction is governed
by S-matrix ideas, have no known Lagrangian. The important scientific legacy
of the bootstrap era is the idea that there are construction methods of QFT
which are not only outside the Lagrangian and functional quantization methods,
but which for the first time are capable to secure the mathematical existence
of models of QFTs.

At the time of the bootstrap Mandelstam \cite{Mand} conjectured a
representation of the elastic scattering amplitude which contained the
dispersion relation as well as an extension of the at that time known momentum
transfer t-analyticity. In contradistinction to the nonlinear bootstrap
program it seemed more susceptible to computational ideas, at least if one
left out unitarity. For the project of establishing the observational validity
of the (model-independent) causality principle underlying QFT via an
experimental check of the dispersion relations, this conjectured but never
proven representation would have little interest; in this case the rigorously
established Jost-Lehmann-Dyson representation clinched the connection between
\textit{causality and the dispersion relation project}.

There was a second more phenomenological motivated train of thought involving
the idea of analytic continuation in the angular momenta and the ensuing
connection of the related Regge trajectories (particles, resonances) which
suggested observational correspondences with the real world of strong
interactions. In this situation Veneziano, guided by the Mandelstam
representation, produced a formula which combined infinitely many particle
poles into a trajectory in such a way that the relation between Mandelstam's s
and t channels appeared as a implementation of the crossing property of QFT;
as a result the model was called the dual model. This was achieved in an
ingenious by mathematical properties of Gamma and Beta functions, so that the
result appeared to some as a profound confirmation of the underlying
phenomenological ideas and to others too mathematical for the phenomenological use.

This construction, which appeared at the beginning to be unique, admitted
several similar solutions \cite{Vech}. As we know nowadays, the correct
interpretation of these somewhat magic constructions (they were not derived or
related to physical principles) has nothing to do with the world of S-matrices
of particle physics. They rather result from the appropriately normalized
\textit{Mellin-transforms of correlation functions of conformal covariant
fields \cite{Mack}. }A conformal 4-point function can be globally
operator-expanded in 3 different ways%
\begin{align}
&  \left\langle A_{1}(x_{1})A_{2}(x_{2})A_{3}(x_{3})A_{4}(x_{4})\right\rangle
\\
&  A_{i}(x)A_{l}(y)=\sum_{k}\int F_{i,j}^{k}(x,y,z)C_{k}(z)d^{4}z\nonumber
\end{align}
The global expansions used in the second line (only) hold in conformal
theories; in general one can only establish the asymptotically convergent
Wilson short distance expansions. As Mack has shown \cite{Mack} there is an
analogy of conformal QFT with particle physics in which the charges of
charge-carrying fields correspond to particle momenta and their anomalous
dimensions to mass squares. This analogy is especially treacherous if one
thinks of the Mellin poles, which lie on the trajectory of the anomalous
dimensions of the composite fields $dimC_{k},$ as the one-particle poles of
the Born approximation of a hypothetical scattering amplitude in a
relativistic QT with infinitely many particles. The three ways of
operator-expanding the conformal 4-point function correspond formally to the
three s,t,u crossed channels. The interpretation as a crossing symmetric Born
approximation in the presence of an infinite tower of interacting particles is
far-fetched, even phenomenological uses of a mathematical formalism cannot
completely ignore its underlying physical principles which point into a quite
different direction; it does however explain the bizarre scientific appearance
which string theory has even to those who do not command over the conceptual
tools used in the present critical evaluation. For more detailed discussion
see \cite{cross}.

One may speculate about what would have happened if this observation would
have arisen in this way and not as it did as a result of experimental
mathematics through tinkering with gamma functions which leaves more leeway
for phenomenological interpretations. Conformal field theory is \textit{the}
field theory par excellence without a particle interpretation\footnote{Any
canonical conformal free field is a free field and the LSZ asymptotic limit of
any anomalous dimensional field vanishes \cite{Sch3} \cite{integrable}. No
inclusive cross section construction as used in QED led to an observable which
could be interpreted in terms of particles.}; its use is limited to the study
of structural problems in QFT. At the time of the invention of the dual model
the above conformal construction was not known. But in plain view of the very
different origin of the particle physics crossing (see next section) a defence
of the dual model/string theory as a description of particle scattering has no
basis. As will be seen in the sequel, the same applies to the use in the sense
of a two-dimensional sigma model with "target-space" indices referring to
higher-dimensional spacetime (the source-target embedding idea)

The mass/spin tower position of the poles in the dual model can also be
obtained from the quantization of a Lagrangian model: the bilinearized
Nambu-Goto Lagrangian\footnote{The original Lagrangian containing square roots
of quadratic form in $X^{\mu}(\sigma,\tau)$ leads to the Pohlmeyer inbariants
which have no relation to string theory \cite{Bahns}.} leads to a wave
function space which contains a tower of irreducible (m,s) Wigner
representation of the Poincar\'{e} group as well as oscillator operators which
link the levels with increasing (m,s). Its second quantization is the
\textit{string field}, an unfortunate name for a dynamic infinite component
pointlike localized field since the name has already been used for fields
causally localized along an semiinfinite spacelike string (unfortunately this
point remained unmentioned in \cite{Dimock}). Here the prefix "dynamic" is
important because a direct sum in the wave function or tensor product in the
second quantized setting would hardly be worthwhile studying; dynamic stands
for the presence of operators which change the composition within the tower
and hence the contribution of the individual components to the
field\footnote{As any field resulting from Lagrangian quantization, the string
field is irreducible (the mathematical meaning of "dynamic") whereas a direct
sum is not.}. In fact the exponential of the field variables $X^{\mu}%
(\sigma.\tau)$ of the model contrary to the belief of string theorists
describe (after splitting off the c .m. motion) a change in the internal
composition over a localization point. In a geometric terminology the change
is upward and not sideways in the target space; it resembles a change of the
spin and not a spread in the pointlike position.

That string theorists have a disturbed relation to causal localization and
covariance is obvious from their supporting use of the classical Lagrangian of
a pointlike particle \cite{Polch}%
\begin{align}
\mathcal{L}  &  =\sqrt{ds^{2}}\curvearrowright x^{\mu}(\tau),~cov.orbit\\
&  no\text{ }quantum~counterpart\nonumber
\end{align}
which is the well-known Lagrangian describing the free movement of a classical
particle in an arbitray $g_{\mu\nu}$ spacetime metric. This one-dimensional
"support" for ST backfires totally since a covariant quantum theory of
relativistic particle orbit does not exist; there is simply no
frame-independent particle position operator a fact which Wigner knew already
in 1939 when he proposed to characterize relativistic
particles\footnote{Wigner was hoping that his relativistic representation
theory would lead to an intrinsic access (without invoking the quantization of
classical structures) to QFT. Whereas the hope was well-founded and found its
later realization in the notion of modular localization, the frame-dependent
Newton-Wigner localization (the Born localization adapted to the relativistic
inner product) was not what he had hoped for.} not by (nonexisting) quantum
mechanical position variables but rather in terms of classifying the
representation spaces of their wave function from which the quantum fields
(which do admit a covariant localization) result by second quantization; its
non-existence is one of strongest reasons for doing QFT instead of
relativistic QM (see a critical evaluation of the "direct particle
interaction" \cite{interface}). There is no covariant QT which originates from
quantization of this Lagrangian and there are many other cases of classical
Lagrangians whose formal quantization does not lead to what one incorrectly
expects, namely a covariant quantum theory. The similar looking classical
Nambu-Goto Lagrangian leads to the same problem; the classical variable has no
quantum counterpart $X_{\mu}(\tau,\sigma)$ which describes a covariant object
in spacetime. Rather what happens with using the (quadratic modification of)
this Lagrangian in ST is that (apart from the zero mode which describes the
covariant position of a pointlike object) the oscillator degrees of freedom in
the Fourier transforms of such objects go into internal degrees of freedom
residing in a Hilbert space which has nothing to do with spacetime
localization. The string field is a pointlike localized field and the only
candidate for a nontrivial \textit{dynamical infinite component field} of the
kind which Barut, Kleinert and others tried to construct it (in vain, becaused
they used extensions of the physical L-group) \cite{Tod}. All correct
calculations of "quantum strings" led to pointlike (graded) commutators, there
are no "points on a string" as in \cite{Lowe} \cite{Martinec}.

Actually such dynamical infinite component fields are very hard to construct
and according to my best knowledge the ST construction is the only one. In
fact Lagrangians which lead to quantum objects with a noncompact internal
symmetry space ("target space") only arise from "second quantization" of
non-rational chiral QFTs since the existence of a continuous charge
superselection structure is the prerequisite of a noncompact target space.
There are simply no higher dimensional quantum field theories with such a
property, a noncompact index structure of such quantum fields must always
comply with the spacetime dimension in which they live (tensor/spinor
indices). A more fundamental theory (QFT) does not have to dance to the tune
of a less fundamental one (classical field theory) i.e. the quantization
parallelism only works with careful selected input; neither do classical
Lagrangians always have quantum counterparts nor do QFT models always have a
classical counterpart. In the case at hand: covariant operators $X_{\mu
}(\sigma,\tau)$ which are associated with a "quantum surface" in spacetime do
not exist (they become "internal") and the claim that ST leads to gravity
cannot be upheld; string theory is a gigantic distorting mirror for particle
physicists albeit a valuable source of intuition for mathematicians.

It should be added that in relativistic QM (DPI) the frame dependence of the
particle position operator does not prevent the large time part of wave
functions to describe covariant trajectories of their c.m., if this would not
be so scattering theory would not even work in QFT. In fact in QFT the
non-covariant localization associated with a particle position operator and
the covariant localization through fields coalesce for infinite timelike
separation of events, and hence the S-matrix matrix is Poincar\'{e} invariant
\cite{interface}. It is this property which in QM allows to talk about the
(material-dependent) velocity of sound in the sense of an effective velocity
even though the wave packets dissipate.

With the exception in chiral QFT (presented below) inner symmetries of scalar
quantum fields, often referred to as sigma-model fields, must transform
according to finite representations of compact groups. This is a consequence
of the profound DHR analysis of superselection rules combined with the DR
construction of field algebras. These theorems use the causal localization of
QFT which is deeply related to vacuum polarization and thermal properties and
as a result is much more restrictive that the geometrical localization of
classical fields. Classical sigma model fields on the other hand may carry
noncompact representations of inner symmetries. This has an obvious
generalization to arbitrary fields: non-scalar quantum fields can, besides
compact internal symmetries only carry the vector/spinor indices which are
associated with the spacetime on which they "live" whereas classical fields
which live on d-dimensional spacetime (the source space) could carry a $D\neq
d$ "target space" symmetry. Apart from the exception of chiral theories such a
source-target embedding is not possible in QFT.

This remarkable exception coming from chiral theories owes its existence to
the fact that there exist besides "rational" chiral models which have a finite
or countable number of superselection sectors also lesser studied models with
a continuous number of inequivalent representation sectors. According to the
DHR superselection theory this cannot happen in higher dimensional theories
\cite{DR}. The simplest such model can be defined in terms of a D-component
current \cite{Longo}\cite{Stas}. Defining potentials $\Phi_{i}$ and using the
weakly convergent integrals over currents which represent the global charges
one has%

\begin{align}
&  \Phi_{i}(u)=\int_{-\infty}^{u}j_{i}(u^{\prime})du^{\prime}%
,~i=1..D,~\left\langle j_{i}(u),j_{k}(u^{\prime})\right\rangle \sim
\frac{\delta_{i,k}}{(u-u^{\prime}+i\varepsilon)^{2}}\label{current}\\
&  \Psi(u,\vec{q})=expi\vec{q}\cdot\vec{\Phi}(x),~Q_{i}\Psi(u,\vec
{q})\left\vert 0\right\rangle =q_{i}\Psi(u,\vec{q})\left\vert 0\right\rangle
,~Q_{i}=\Phi_{i}(\infty)\nonumber
\end{align}

\begin{align}
&  Q_{i}\rightarrow P_{i}~~multcomp.~charge~q_{i}\sim
particle~momenta~p~~\label{ana}\\
&  \vec{q}^{2}=dim\Psi,~particle\text{ }mass~p^{2}~\sim anomalous\text{
}dim.\nonumber
\end{align}

The last two lines contain an analogy between the charge- and momentum
labeling but it needs to be emphasized that at this point this is only
suggestive. This model has a rich use in mathematical physics. The general
commutation relations between the $\Psi$ are plektonic (braid group) and one
of the interesting problems has been to classify the maximal local extension
of the vacuum representation of the $j_{i}$ algebra.$~$This can be done in
terms of even lattices. Among those maximal observable algebras there are some
which have no superselection sectors and belong to the largest exceptional
finite groups (the moonshine group).

The use of this chiral sigma model for particle physics consists in trying to
convert the above analogy into a situation where the D-dimensional space
carries a representation of the Poincare group. This requires a Fourier
decomposition of the compactified conformal field $\Phi_{i}$ and the use of
the zero mode as the coordinate $x_{i}$ whereas the higher oscillator modes
describe the changing internal degrees of freedom and not a movement in target
spacetime. With other words the sigma field remains point-like
localized\footnote{It is a solution of the old problem of finding dynamical
infinite component fields for which prior attempts to generate an
representation containing an interesting infinite (m,s) tower spectrum failed
\cite{Oksak}.} since the infinitely many oscillator degrees of freedom build
up an internal space over the localization point. In \cite{Martinec}
\cite{Lowe} the authors computed the commutator of the infinite component
string field and obtained the correct result of its pointlike nature. But
(perhaps by being a member of the string community) they overlook this
important point and rather create the impression that one is confronting a
string of which only one point is visible. Trying to please a community or self-delusion?

If the Poincar\'{e} group representation is required to be unitary and with
positive energy then the answer is almost unique\footnote{The requirement that
a unitary positive energy representation of the Poincar\'{e} group acts on the
inner symmetry target space of a non-rational chiral sigma model determines a
mass/spin spectrum.}: it must be the D=10 super string representation
respectively one of its finitely many M-theoretic variations. Such a nearly
unique answer is always surprising; no representation at all or infinitely
many would have attracted less attentions. Nevertheless it is hard to
understand the immense physical-philosophical leap from the observation that
non-rational chiral sigma models can carry representations of noncompact
groups (in particular of the Poincar\'{e} group) to the invocation of a
fundamental theory of our living spacetime.

Even in this case of the use of non-rational chiral models the representation
on the target space of continuous superselection sectors the target
localization remains pointlike, there is no embedding of a chiral theory in a
higher dimensional target space; in fact a lower dimensional QFT can never be
embedded into a higher dimensional target theory, QFT is too
holistic\footnote{Holistic refers to the fact that that localization in QFT
is, different from QM since it is always accompanied by thermal manifestations
and vacuum polarization at the causal horizon of the spacetime localization
region. In \cite{hol} and \cite{cross} the reader finds a more detailed
presentation and illustrations of this concepts.} in order to allow such a
possibility; for an explanation of "holistic" see \cite{hol} \cite{cross}. A
string-like localized object cannot be introduces in terms of embedding; the
only known way to obtain a nontrivial\footnote{One which is not representable
as a line integral over a pointlike field.} string-like localized structure in
a higher dimensional spacetime is to introduce an elementary
\textit{stringlike localized field}. An example for a covariant stringlike
free field is provided by the infinite spin Wigner representations \cite{MSY}.
These fields have no Lagrangians in contrast to the fields of string theory.
So this state of affairs may be paraphrased by: string theory is Lagrangian
but not string-localized and string-localized fields are genuinely
string-localized but do not arise from Lagrangian quantization. The Nambu-Goto
and the Polyakov Lagrangian are classical strings but their canonical
quantization do not produce \textit{quantum world sheets} inasmuch as the
relativistic particle Lagrangian does not produce a \textit{quantum world
line}. If one wants to associate a quantum theory with the N-G Lagrangian it
is the theory of Pohlmeyer's invariants \cite{Bahns}. They are the quantum
counterparts of classical invariants associated with the N-G Lagrangian, but
their quantum interpretation is unknown.

The above critique applies to the embedding interpretation of objects $X^{\mu
}(\sigma,\tau)$ $\mu=1..D$ in functional integrals as world sheets in a
D-dimensional target space. One has no control over the localization of such
objects and it should therefore be no surprise that these degrees of freedom
go, apart from a zero mode which goes into the pointlike localization, into
the internal degrees of freedom over a point. The holistic structure of QFT
forbids any form of embedding. This affects also Polyakov's Lagrangian
formulation of string theory and its use as a basis of gravity. It also
affects the quantum counterpart of the Kaluza-Klein dimensional reduction. A
QFT characterized by correlation functions with an analytic continuation to
euclidean points can be dimensionally reduced by "thermalization" in one
euclidean coordinate and subsequently taking the infinite temperature limit
which shrinks the periodicity $\beta\rightarrow0$ and may be interpreted as a
"curling up" one coordinate. Granting sufficient analyticity this construction
via thermal states maintains the original algebraic structure and remains
therefore intrinsic. On the other hand, Kaluza-Klein arguments on the level of
the Lagrangians instead of the quantum correlation functions of a model
\cite{Wi} are not trustworthy since they ignore the holistic structure of QFT
correlation functions.

A related mechanism to lower the spacetime dimension of a QFT is the
holographic correspondence/projection. The best known case is the
$AdS_{d+1}\longleftrightarrow CFT_{d}$ correspondence which relates models on
d+1 dimensional Anti-de-Sitter space with conformal fields. Whereas the
descend from the higher to the lower dimensional theory can still be
formulated in terms of spacetime limits of AdS fields, the inverse map is more
conveniently expressed in an algebraic setting \cite{Re} since there is no
pointlike correspondence. Although this correspondence can be rigorously
established on the mathematical side, it suffers from a serious physical
defect which shows up if one asks questions about the cardinality of phase
space degrees of freedom. Intuitively one would expect to find an
"overpopulation" on $CFT_{d}$ if one starts from a "healthy" QFT on
$AdS_{d+1}$ and an "anemic" $AdS_{d+1}~$theory in the opposite case.

This is indeed the case; although both theories are Einstein-causal, the too
many degrees of freedom in the lower dimensional theory lead to a breakdown of
the causal shadow property \ref{c}. This can be made explicit by computing the
result starting with a free massive field on AdS in which case the
correspondence leads to a conformal generalized free field, an object which
violates this aspect of the causality property and for who's exclusion the
causal shadow property was introduces in \cite{H-S}. The rational for this at
that time was that although it is desirable to dissociate QFT from the
quantization parallelism to classical field theory, one should preserve all
those properties which are formally part of the Lagrangian formalism (the
causal Cauchy propagation) but which allow a formulation within a relativistic
QT. In a later work \cite{Ha-Sw} this was connected with the cardinality of
phase-space degrees of freedom which in QFT mildly infinite (compact, nuclear)
\cite{Haag} in contrast to QM where it is finite per cell of phase space.

Another closely related variation of the same theme is the conceptual
interpretation of the restriction of conformal QFT to a lower-dimensional
"conformal brane" . In this case the too many degrees of freedom can be
encoded into a huge internal symmetry \cite{Mack} which however does not
change the unphysical aspect. Since D-branes have only been proposed as
quasiclassical objects, the issue of degrees of freedom remained unaddressed.

Phase space arguments were not really popular in the old days because QFT
outside Lagrangian quantization was not considered a pressing issue; in more
recent times, when the degree of freedom issue could have had a moderating
influence on an issue (on which part of QFT came apart at its seams in the
flurry of the Maldacena conjecture), it was largely forgotten.

Since our conclusions about all problems in which causal localization come
into play were negative, let us try to understand whether at least the
retraction to a pure S-matrix setting, forgetting Lagrangians and string
fields, is consistent. Here it is instructive to take a critical look at the
first attempt by Heisenberg and somewhat later by Stueckelberg to formulate
particle physics in a pure S-matrix setting. As mentioned in the introduction
Stueckelberg added the idea of macrocausality to Heisenberg's Ansatz in terms
of an exponential phase operator which only satisfied the
cluster-factorization of S. This requirement led him to the asymptotic
structure of causal rescattering in terms of particle lines representing
Feynman propagators. By extrapolating this structure for all distances and
assuming that the interaction region is pointlike, he found the Feynman rules,
i.e. by the extension of the macrocausality property for particles beyond
their range of validity, he arrived at the perturbative rules of micro-causal QFT.

It is interesting to mention that Stueckelberg's approach \cite{Stue} actually
amounts to an iterative unitarization process for an S-matrix, starting with
the first order%
\begin{align}
S  &  =1+\sum_{k=1}^{\infty}S_{k},~~S_{1}+S_{1}^{\ast}=0\\
S_{2}+S_{2}^{\ast}  &  =-S_{1}S_{1}^{\ast},~S_{3}+S_{3}^{\ast}=-(S_{1}%
S_{2}^{\ast}+S_{2}S_{1}^{\ast}),~etc\nonumber
\end{align}
He started (in modern terminology) from a anti-Hermitian S$_{1}$ in terms of a
spacetime integral over a scalar Wick-ordered polynomial in free fields. Since
in each order only the Hermitian part is determined by the previous orders, he
imposed micro-causality, even though in contrast to macro-causality this had
no cogent reason in a particle S-matrix setting and requires to write S$_{1}$
in terms of pointlike free fields. He then showed that this restricted
unitarization scheme determines S up to local counterterms and the result is
then identical to Feynman's time-ordered formalism and therefore imposing the
micro-causality on the operator densities of the various $S_{k}$ is equivalent
to perturbative QFT \footnote{The iteratively constructed S(g) is the
Stueckelberg-Bogoliubov-Shirkov operator functional which depends on
space-time dependent coupling function. The S-matrix (formally for constant
g's) is related to this functional by the "adiabatic limit" whose existence is
roughly equivalent to the asymptotic convergence of fields in the LSZ
scattering theory, In many theories (eg. QED) this limit does not exist as a
result of the infraparticle phenomenon \cite{charge}.
\par
\bigskip
\par
\bigskip
\par
{}}.

Forgetting for a moment that the dual model is based on the Mellin transform
crossing property rather than the particle physics crossing (next subsection)
let us ask the question: what does it mean to obtain the $S_{scat}$ of string
theory by unitarizing the dual model? Where is the first order operator in a
Hilbert space which corresponds to the string S-matrix analog of
Stueckelberg's $S_{1}$? And what controls the higher order anti-Hermitian
parts? And last not least what does the notion "stringy" with its spacetime
appeal mean, assuming that the stringyness of an S-matrix (whatever it means)
entered the theory through the dual model. Does it imply reading back an
S-matrix property into an imagined off-shell property? Why should calculations
of functions called scattering functions by fiat, based on splitting and
fusing tube graphs (which for 50 years resisted their presentation in terms of
operators and states) have anything to do with QT?

In the next subsection it will be shown that despite all these failures of
S-matrix based approach to particle physics the S-matrix remains the most
important tool in a non-perturbative approach to particle theory. This new
approach will also lead to a derivation of the particle crossing property and
lend new strength to the belief that the S-matrix is not only the roof of QFT
but also an indispensable tool in the nonperturbative classification and
construction of its models.

\subsection{ Modular localization, its thermal manifestation and the origin of
particle crossing}

\textsc{In diesem Fall und ueberhaupt, kommt es ganz anders als man glaubt.
(W. Busch)}

Many properties in QFT allow a more profound understanding (beyond the mere
descriptive presentation) in a formulation in which one deals with space-time
indexed systems of operator algebras rather than with their generating point-
or string-like localized fields. An illustration was given before within the
algebraic formulation of causality, in which the \textit{causal shadow
property} is more natural then its formulation in terms of quantized Cauchy
data for individual fields. This is in particular true about the thermal
properties which result from the restriction of the vacuum (or other finite
energy states) to local subalgebras.

This thermal manifestation of subalgebras is a holistic property par
excellence; although the KMS which characterizes this manifestation is shared
by all individual operators which share the same causally completed
localization region, its conceptual understanding is only possible in terms of
ensembles of observables; just as in the global standard "heat bath"
statistical mechanics. Whereas in QM such situation arises mainly through a
coupling to heat bath within the thermodynamic limit of Gibbs density
matrices, in QFT the restriction of the vacuum (or any other finite energy
state) to a localized algebra creates a singular impure KMS state without the
coupling to an external heat bath.

The appearance of "localization thermality" in QFT (in contrast to the global
heat bath caused thermal aspects of thermodynamics) has far-reaching
consequences. In a way it vindicates Einstein's steadfast rejection of Born's
probability as an attribute of an individual quantum mechanical event. In his
dispute with Jordan\footnote{Nowadays referred to as the Einstein-Jordan
conundrum \cite{Du-Ja}. It played an important role in Heisenberg's later
discovery of vacuum-polarization, but the understanding of localization
thermality had to wait more than 6 decades.} \cite{Ei-Jo} both of them missed
to notice this important conceptual attribute of Jordan's new "wave
quantization". Its early recognition could have changed the history of the
philosophy of QT (in particular concerning the measurement process) and
reconciled Einstein with its probabilistic ensemble structure.

A property which comes directly from the adaptation of the "action at the
neighborhood principle" of Faraday and Maxwell to QT \cite{Lille} to QT and
does not have to be added "by hand" (as Born's quantum mechanical
probability\footnote{The localization probablity of a Schr\"{o}dinger wave
function and its relation to the spectral decomposition of the localization
operator was introduced soon after Born defind the scattering probability
(cross section) for the Born approximation. by Pauli; it appears as an added
footnote in Born's paper.}) would almost certainly have received Einstein's
approval. In QFT the basic observables are localized; quantum mechanical
global observables and the assignment of probabilities to individual events as
a \textit{practically useful idealization} (and not as a foundational
property) hardly cause philosophical headaches. It is regrettable that
philosophers of science have remained with Schr\"{o}dinger's cats and
Everett's many-world interpretation; the messages coming from the less
metaphoric and more fundamental local quantum physics \cite{Haag} have largely
been ignored. In the early days the holistic aspects of QFT as manifested in
vacuum polarization and thermal aspects of localization were not understood
since QFT was thought of in terms of relativistic QM \cite{interface}, but
there is no reason why even nowadays questions of interpretation of QT are
mainly discussed in the setting of the less fundamental QM.

In the following we will present some important results from local quantum
physics \cite{Haag}; the reader who is unfamiliar with these concepts should
consult the literature \cite{interface}.

Spatial separation of a quantum mechanical global algebra into two mutually
commuting spatially separated (inside/outside) subalgebras leads to a
\textit{tensor factorization} of the quantum mechanical vacuum state and the
phenomenon of \textit{entanglement} for general states. Nothing like this
happens for local subalgebras in QFT. Rather they radically change their
algebraic properties; instead of being equal to the algebra of all bounded
operators on a smaller Hilbert $B(H_{red})$ space referring to the spatially
reduced degrees of freedom as in QM, the localized subalgebras of QFT belong
to a different type called "hyperfinite type III$_{1}$ factor algebra" (in the
Murray-von Neumann-Connes classification) for which among several other
changes the above tensor-factorization breaks down. For reasons which will
become clear later on, we will call this operator algebra type shortly a
\textit{monad,} so every localized algebra of QFT is a \textit{monad}
\cite{interface}. Its occurrence in QFT is inseparably related to the vacuum
polarization and thermal properties of localization in QFT.

Although the division into a spacetime region and its causal complement leads
by use of Einstein causality to the mutual commutativity, the \textit{vacuum
does not factorize} and hence the prerequisites for the usual form of
entanglement are not fulfilled; this is in spite of the fact that the algebra
associated to a spacetime region and its commutant together generate the full
global algebra $B(H)$. In fact \textit{a monad has no pure states} at all,
rather all states are \textit{impure in a very singular way}, i.e. they are
not density matrix states as impure states in QM; the vacuum restricted to a
local monad turns into a singular KMS state. Such states appear in QM only in
the thermodynamic limit of Gibbs states on box-quantized operator algebra
systems in a volume $V$. Later we will turn to the mathematics behind these
observations which is \textit{modular operator theory} and more specific
\textit{modular localization}.

Usually people are not interested\footnote{For example the tensor
factorization formalism known as "thermo-field formalism" breaks down in the
thermodynamic limit for the same reasons as the Gibbs density matrix
description, i.e. this formalismus is not suited to describe "open systems".}
in an intrinsic description of the thermodynamic limit state (called
"statistical mechanics of open systems"), but if they were, they would find,
as Haag, Hugenholtz and Winnink in 1967 \cite{Haag}, that the limiting state
ceases to be a density matrix state and becomes instead a singular KMS state
i.e. a state which has lost its trace property and hence its Gibbs
representation property (volume divergence of partition function). Instead it
fulfills an analytic relation which first appeared as a computational trick
(to avoid computing traces) in the work of Kubo, Martin and Schwinger and
later took on its more fundamental significance \cite{Haag} which it enjoys
presently as one of the most impressive links between physics and mathematics.
Whereas monads with singular KMS states appear in QM \textit{only} in the
thermodynamic limit of finite temperature Gibbs states, their occurrence in
QFT is abundant since any reduction of the vacuum onto a causally closed
localized subregion leads to such an impure state which is KMS "thermal" with
respect to the modular Hamiltonian on a monad (see below).

With a "split" of size $\varepsilon$ between the subalgebra and its causal
complement\footnote{The possibility of doing this is called "the split
property" \cite{Do-Lo}. Wheres the standard box quantization does not allow to
view the boxed system as a subsystem of a system in a larger spacetime, the
splitting achieves precisely this at the price of vacuum polarization at the
boundary. The physics based on splitting is called "open system" setting.},
one returns to a situation which resembles QM in that the global algebra
tensor-factorizes. The possibility of doing this is called "the split
property" \cite{Do-Lo} and it is closely related to issue of what replaces the
quantum mechanical finite phase space degrees of freedom per unit phase space
cell in QFT \cite{Haag} and the closely related question of the causal
propagation as expressed in the time-slice property and "Haag duality".
Whereas the standard quantum mechanical box quantization does not allow to
view the boxed system as a subsystem of a system in a larger spacetime, the
split property achieves precisely this at the price of vacuum polarization at
the boundary. The physics based on splitting is called the "open system"
setting \cite{Haag}.

The state in which the system "splits" is a highly entangled Gibbs density
matrix state with respect to a "split Hamiltonian" which approximates for
$\varepsilon\rightarrow0~$the modular Hamiltonian (see below) which is
determined by the localized algebra and the vacuum state ($\mathcal{A(O}%
),\Omega$). There are presently two pictures leading to two formulas which are
different by a logarithmic factor: the lightlike sheet picture and the analogy
with Heisenberg's partial charge behavior. The latter leads to (n=spacetime
dimension)%
\begin{equation}
entropy\sim%
\genfrac{\{}{.}{0pt}{}{ln(\frac{R}{\Delta R}),~n=2}{(\frac{R}{\Delta R}%
)^{n-2},~n>2}%
\end{equation}
where the region is a ball of radius $R~$and~$\Delta R~$the "fuzziness" of its
boundary (i.e. $\varepsilon\symbol{126}\frac{\Delta R}{R}$). This is identical
to the behavior of Heisenberg's partial charge $\left\Vert Q_{R,\Delta
R}\Omega\right\Vert $ acting on the vacuum. This is result which one can
rigorously derive in terms of appropriate test function smearing from the
two-point function of a conserved current \cite{E-J}. This plain area law also
follows from 't Hooft's brickwall picture \cite{Ho}.

On the other hand the light sheet picture favours a logarithmic modification
of the area law for n%
$>$%
2. The picture is that of a volume factor of a "box" where two transverse
spatial direction are responsible for an area factor and the third lightlike
extension contributes a $ln(\frac{R}{\Delta R})$ factor which substitutes the
factor from the third spatial extension \cite{foun}\cite{BMS}. It is most
close to the idea of an inverse Unruh effect since the space box of heat bath
statistical mechanics is replaced by a box for which one direction is
lightlike (being responsible for the log modification\footnote{In \cite{E-J}
it was called the \textit{weak Unruh inverse;~}the terminology \textit{strong
Unruh inverse} being reserved for an \textit{isomorphism} between a heat bath-
and a localization caused- thermal system.}). The abstract nature of the split
property and the difficulties in extracting a concrete computation from it
(the approximating Hamiltonian is not known) make it presently impossible to
decide between a plain area law and its logarithmic modification. This
relation between the sharpness $\varepsilon$ of the localization boundary and
the localization entropy/energy replaces the uncertainty relation of QM
whereas Heisenberg's uncertainty relation becomes meaningless since there is
no position operator in QFT.

The connection between modular localization and thermal aspects may be little
known, but there is one system which made it into science fiction and is part
of the particle physics folklore: the aforementioned Unruh Gedankenexperiment,
which is nearly as old as the close related Hawking effect. In both cases
quantum fields become localized, in the first case behind the
observer-dependent Rindler wedge which is the causal shadow region of its
horizon (i.e. the wedge itself), and in the second case the less "fleeting"
(less observer dependent) situation of the \textit{event horizon} of the
Schwarzschild metric. In the first case the important question, which Unruh
answered by the construction of a Gedankenexperiment, was \textit{what is the
physical meaning of being localized in a wedge W of Minkowski spacetime}? In
this case the modular Hamiltonian is the generator of the W-preserving Lorentz
boost i.e. the W-localized observable (counter, observer) must be uniformly
accelerated in order not to trespass the horizon of the wedge; for him the
inertial frame Hamiltonian of Minkowski spacetime is irrelevant, his
Hamiltonian is the spectral-symmetric boost generator (instead of the one
sided spectrum of the time translation Hamiltonian). This requires to pump
energy to accelerate an observable i.e. the Unruh effect is not an "perpetuum
mobile" for creating heat, and the vacuum on the global algebra of all
operators in Minkowski spacetime keeps its ground state properties with
respect to the Poincar\'{e} group.

In order to remove one more mystery from the connection of localization with
the thermal aspects of the reduced vacuum and the concomitant effect of vacuum
polarization, we will now show how the important \textit{crossing property} of
particle physics has its explanation in the KMS property of the
wedge-restricted vacuum. Although the mystery which surrounded the crossing
property for many decades and led to incorrect interpretations (see previous
section) will be lifted, some surprise about its unexpected true nature
remains; this is what the line added to the title of this section refers to.

For this purpose one starts from the modular operator theory applied the wedge
algebra $\mathcal{A}(W)$ which denotes the operator algebra formally generated
by smeared fields whose smearing function support is contained in
$suppf\subset W.$ For this algebra acting on the vacuum $\Omega$ the Tomita
S-operator is well-known to have the following definition and lead to
important operators under polar decomposition\cite{Haag}%
\begin{align}
&  ~S_{W}A\Omega=A^{\ast}\Omega,~~A\in\mathcal{A}(W),~~S_{W}=J_{W}\Delta
_{W}^{\frac{1}{2}}\label{modular}\\
&  \Delta_{W}^{i\tau}=U(\Lambda(-2\pi\tau)),~J_{W}=S_{scat}J_{W,0}%
,~S_{W,0}=J_{W,0}\Delta_{W}^{\frac{1}{2}}\nonumber
\end{align}
Here the modular unitary$~\Delta_{W}^{it}$ is shared between the interacting
and free (incoming) system which carries the additional subscript $0$. The
antiunitary $J_{W,0}$ which appear in the polar decomposition of $S_{W,0}$
represents the reflection on the edge of the wedge W (TCP, apart from a $\pi
$-rotation along the wedge) without interaction whereas $J_{W}$ includes the
interaction in form of the S-matrix $S_{scat},$ which now plays an additional
role to the one in scattering theory, namely that of a \textit{modular
invariant} between the free (incoming) and the interacting wedge algebra
\cite{interface}.

The equality of the dense domains of the interacting $S$ with that of the free
$S_{0}$ i.e. $domS=domS_{0}=dom\Delta^{\frac{1}{2}}$ implies that there is a
dense set of states, namely those in $dom\Delta^{\frac{1}{2}}$ which can be
generated both in the interaction free algebra $\mathcal{A}_{in}(W)$ generated
by the W-smeared incoming fields and operators from the interacting algebra
$\mathcal{A}(W).~$Hence for each operator $A\in\mathcal{A}_{in}(W)$ there
exists a $A_{\mathcal{A}(W)}\in\mathcal{A}(W)$ such that
\begin{equation}
A\left\vert 0\right\rangle =A_{\mathcal{A}(W)}\left\vert 0\right\rangle
\label{bij}%
\end{equation}
The definition is well-defined and bijective because the vacuum is separating
for both algebras $\mathcal{A}_{in}(W)~$and $\mathcal{A}(W)$ and because the
dense subspaces $\mathcal{A}_{in}(W)\left\vert 0\right\rangle $ and
$\mathcal{A}(W)\left\vert 0\right\rangle $ coincide, both being equal to the
domain of the common modular operator $\Delta^{\frac{1}{2}}.$ We shall refer
to the bijective assignment
\begin{equation}
\mathcal{A}_{in}(W)\ni A\rightarrow A_{\mathcal{A}(W)}\in\mathcal{A}(W)
\end{equation}
as "emulation" (of a interaction-free operator $A$ in the interacting algebra
$\mathcal{A}(W)$). The emulation of a single particle operator was previously
called PFG in \cite{BBS}, hence a PFG is the emulat of a smeared free field
$A(f)$ with $suppf\subset W$ and a Wick-product $:A(f_{1})..A(f_{n}):$ leads
to a multi-particle emulat $:A(f_{1})..A(f_{n}):_{\mathcal{A}(W)}$

The uniqueness of the emulats is secured by demanding that the dense domain
$\mathcal{A}(W)^{\prime}\left\vert 0\right\rangle $ contains a core of the
emulats. From the definition it is clear that emulation is not an algebra
homomorphism $(AB)_{\mathcal{A}(W)}\neq A_{\mathcal{A}(W)}B_{\mathcal{A}(W)}$
and $\left(  A^{\ast}\right)  _{\mathcal{A}(W)}\neq\left(  A_{\mathcal{A}%
(W)}\right)  ^{\ast}.$ More precisely, it follows that%
\begin{equation}
\left(  A_{\mathcal{A}(W)}\right)  ^{\ast}\left\vert 0\right\rangle
=SA_{\mathcal{A}(W)}\left\vert 0\right\rangle =S_{scat}S_{0}A\left\vert
0\right\rangle =S_{scat}A^{\ast}\left\vert 0\right\rangle =S_{scat}A^{\ast
}S_{scat}^{-1}\left\vert 0\right\rangle \label{star}%
\end{equation}

With the help of these emulats one can now state the cyclic KMS property for
the interacting algebra in a form which is convenient for the later derivation
of particle crossing%

\begin{align}
&  \left\langle 0|BA_{\mathcal{A}(W)}^{(1)}A_{\mathcal{A}(W)}^{(2)}%
|0\right\rangle =lim_{\tau\rightarrow1}\left\langle 0|A_{\mathcal{A}(W)}%
^{(2)}\Delta^{\tau}BA_{\mathcal{A}(W)}^{(1)}|0\right\rangle \label{K}\\
&  A^{(1)}\equiv:A(f_{1})...A(f_{k}):,~A^{(2)}\equiv:A(f_{k+1})...A(f_{n}%
):,~suppf_{i}\in W\nonumber
\end{align}
The analytic content of the crossing relation is that the right hand side is
analytic in $0<\tau<1$ and that the analytic continuation of the right hand
side from the physical value $\tau=0$ to $\tau=1$ ($\mathbf{1}$ to $\Delta$)
obeys the crossing identity (\ref{K}). The second line specifies the
$W$-localized Wick products ($suppf_{i}\subset W$) which acting on the vacuum
convert the right hand side into a formfactor of $B$ (or formally $\Delta B)$
between a k-particle incoming bra- and a n-k particle \textit{outgoing} (since
the star (\ref{star}) involves the S-matrix) ket state.

In the absence of interactions i.e. with B being a Wick composite of a free
field, the content of the free KMS relation is a rather direct consequence of
the Wick-ordering theorem. The particle content of the identity%
\begin{align}
&  \left\langle 0|BA_{\mathcal{A}(W)}^{(1)}|\hat{f}_{k+1},..,\hat{f}%
_{n}\right\rangle _{in}=~_{out}\left\langle \hat{f}_{a,k+1},..\hat{f}%
_{a,n}|\Delta B|\hat{f}_{1},..,\hat{f}_{k}\right\rangle _{in}=\label{id}\\
&  =\int\frac{d^{3}p_{1}}{2p_{0,1}}..\int\frac{d^{3}p_{n}}{2p_{0,n}}\hat
{f}_{1}(p_{1})..\hat{f}_{n}(p_{n})~_{out}\left\langle -\bar{p}_{k+1}%
,..-\bar{p}_{n}\left\vert \Delta^{\frac{1}{2}}B\right\vert p_{1}%
,..p_{k}\right\rangle _{in}\nonumber\\
&  _{out}\left\langle -\bar{p}_{k+1},..-\bar{p}_{n}\left\vert B\right\vert
p_{1},..p_{k}\right\rangle _{in}:=~_{out}\left\langle \bar{p}_{k+1},..\bar
{p}_{n}\left\vert \Delta^{\frac{1}{2}}B\right\vert p_{1},..p_{k}\right\rangle
_{in}\nonumber
\end{align}
is however highly subtle; the culprit is the uniquely determined but not yet
known operator $A_{\mathcal{A}(W)}^{(1)},$ which still needs to be computed.
The first line uses (\ref{star}) which converts the action of the conjugate of
$A_{\mathcal{A}(W)}^{(2)}$ on the bra-vacuum into an outgoing particle state
vector of antiparticle $\bar{p}$ in antiparticle wave functions $\hat{f}_{a}.$
Their complex conjugate outside the matrixelement is then transformed back
into particle wave functions by using $\Delta^{\frac{1}{2}}$ of the $\Delta$
for its analytic continuation back to the $\hat{f}$ (last two lines in
(\ref{id}))$.$ If we could forget the emulation in the middle of the left hand
side and equate the resulting integrands in momentum space, we would obtain
the "folkloric" version of the crossing identity (which turns out to be only
valid for special configurations), but the correct version is much more subtle
and interesting.

Let us first be reminded how this problem was resolved for integrable d=1+1
models in \cite{BKFZ} for which the plane wave particle states can be uniquely
described in terms of rapidities $\theta.$ The crucial step was to think about
analytic transpositions i.e. to consider the vacuum formfactor of
two-dimensional models in the rapidity parametrization
\begin{equation}
\left\langle 0\left\vert B\right\vert \theta_{1},\theta_{2},..\theta
_{n}\right\rangle _{in}%
\end{equation}
as a locally analytic functions in the $\theta s$ and that by analytic
continuation we can change the order of $\theta s.~$In that case one arrives
at new objects and in order to not confuse these new objects with the trivial
result obtained from statistics we, stipulate (using the statistics
degeneracy) that for the natural (from left to right decreasing) ordering
$\theta_{1}>..>\theta_{n}$ the state in the vacuum formfactor refers to an
incoming n-particle state and any other left right ordering denotes a yet
unknown object which is defined by analytically continuing starting from the
natural order to that obtained by analytic continuation.

In \cite{BKFZ} it was shown that for models whose S-matrix is given in terms
of an elastic two-particle scattering function $S^{(2)}(\theta_{1}-\theta
_{2})$ (or a matrix of scattering functions), the analytic transposition of
two adjacent $\theta$ is described by this scattering function. In this case
the S-matrix as well as the resulting formfactors are meromorphic functions on
the multi-$\theta$ plane. The analytic change of the ordering can be encoded
into a wedge localized operator whose positive and negative energy components
fulfill the Zamolodchikov-Faddeev algebra relation%

\begin{align}
\left(  \tilde{A}_{in}(x\right)  _{\mathcal{A}(W)}  &  =\int_{\partial
C}Z^{\ast}(\theta)e^{ip(\theta)x}d\theta,~C=(0,i\pi)~strip~\label{ZF}\\
Z(\theta)Z^{\ast}(\theta^{\prime})  &  =\delta(\theta-\theta^{\prime
})+S(\theta-\theta^{\prime}+i\pi)Z(\theta^{\prime})Z(\theta),~Z(\theta
)=Z^{\ast}(\theta+i\pi)\nonumber\\
~Z^{\ast}(\theta)Z^{\ast}(\theta^{\prime})  &  =S(\theta-\theta^{\prime
})Z^{\ast}(\theta^{\prime})Z^{\ast}(\theta)\nonumber
\end{align}
and the product in the natural order%
\begin{align}
Z^{\ast}(\theta_{1})..Z^{\ast}(\theta_{n})\left\vert 0\right\rangle  &
=\left\vert \theta_{1},..\theta_{n}\right\rangle _{in}\label{in}\\
Z^{\ast}(\theta_{n})..Z^{\ast}(\theta_{1})\left\vert 0\right\rangle  &
=\left\vert \theta_{1},..\theta_{n}\right\rangle _{out} \label{out}%
\end{align}

Using the product representation of the n-particle S-matrix in terms of
$S^{(2)}$ it is easy to see that the opposite order corresponds to the
outgoing n-particle state (\ref{out}). The algebraic encoding of the analytic
changes in terms of positioning of noncommutative operators instead of
$\theta$-ordering in n-particle state vectors is evidently more convenient.
\ In particular one obtains for the action of a $Z^{\ast}$\bigskip
$(\vartheta)$ on an ordered n-particle state a n+1 particle state multiplied
with a numerical factor%

\begin{align}
Z^{\ast}(\vartheta)\left\vert \theta_{1},.\theta_{n}\right\rangle _{in}  &
=S_{gs}^{\left(  j\right)  }(\vartheta;\theta_{1},.\theta_{j})\left\vert
\theta_{1},.\theta_{j},\vartheta,\theta_{j+1},..\theta_{n}\right\rangle
_{in},~~\theta_{j}<\vartheta<\theta_{j+1}\label{Z}\\
S_{gs}^{\left(  j\right)  }(\vartheta;\theta_{1},.\theta_{j})  &
=S(\vartheta-\theta_{1})..S(\vartheta-\theta_{j})=S^{\ast}(\theta_{1}%
,..\theta_{j})S(\vartheta,\theta_{1}..\theta_{j}) \label{ord}%
\end{align}
where we call $S_{gs}^{\left(  j\right)  }(\vartheta;\theta_{1},.\theta_{j})$
the \textit{grazing shot S-matrix} for $\vartheta$ impinging on the
$\theta_{1},..\theta_{n}$ cluster\footnote{Note that as a result of momenta
conservation the dull S-matrices as well as their grazing shot counterparts
for integrable models allow a simpler notation. The matrixelements of the
n-particle S-matrix is a combinatorial product of two-particle amplitudes
\cite{BKFZ}.}. For the generalization to the non-integrable case it is
important to express $S_{gs}$ in terms of a product of ordinary S-matrices
(\ref{ord}). For the action of the annihilation operator one obtains a grazing
shot S-matrix with $\vartheta\rightarrow\vartheta+i\pi$ which multiplies an
n-1 particle state since once the $Z(\vartheta)$ has been commuted through the
cluster it annihilates the $\theta_{j+1}$ (the $\check{\theta}_{j+1}$
indicated that it is missing)%
\begin{equation}
Z(\vartheta)\left\vert \theta_{1},.\theta_{n}\right\rangle _{in}%
=\delta(\vartheta-\theta_{j+1})S_{gs}^{\left(  j\right)  }(\vartheta
+i\pi;\theta_{1},.\theta_{j})\left\vert \theta_{1},..\check{\theta}%
_{j+1},..\theta_{n}\right\rangle _{in}%
\end{equation}
The grazing shot scattering amplitude of a $\vartheta$-"bullet" impinging on a
$\theta$-cluster has a generalization to non-integrable QFT%

\begin{align}
S_{gs}^{(m,n)}(\vartheta,j;\chi,\theta)  &  \equiv\sum_{l}\int.\int
d\vartheta_{1}..d\vartheta_{m}\left\langle \chi_{1}..\chi_{m}|S^{\ast
}|\vartheta_{1},.\vartheta_{l}\right\rangle \cdot\label{gs}\\
&  \cdot\left\langle \vartheta,\vartheta_{1},.\vartheta_{l}|S|\vartheta
,\theta,\theta_{1},.\theta_{j}\right\rangle \nonumber
\end{align}
We conjecture that this expression describes the analog of (\ref{Z}) i.e.%
\begin{equation}
_{in}\left\langle \chi_{1},.\chi_{m}\right\vert Z^{\ast}(\vartheta)\left\vert
\theta_{1},.\theta_{n}\right\rangle _{in}=S_{gs}^{(m,n)}(\vartheta
,j;\chi,\theta) \label{bil}%
\end{equation}
Note that in this case the re-ordering into the natural order brings in
particle state vectors of arbitrary high particle number. The intuitive idea
behind the grazing shot S-matrix is that commuting a PFG through a cluster of
particles, the result should be trivial if the localization of the wave packet
of the particle remains effectively in a large distance from that of the
cluster. Such $Z^{\#}(\vartheta)s$ cannot characterized in terms of Z-F like
commutation relations.

The argument leading to a the action of a one-particle emulat $Z^{\#}$ on an
incoming n-particle state is the same in both cases. It rests on two assumption:

\begin{enumerate}
\item The only way the interaction enters this action is through the S-matrix

\item The action is determined with the help of $\theta$-reordering defined in
terms of commuting the emulat $Z^{\#}$ through $\theta$-clusters.
\end{enumerate}

The first assumption receives its support from the fact that the interaction
enters the modular object $J=S_{scat}J_{in}$ only through the scattering
matrix, hence one expects that the generating emulats of $\mathcal{A(}W)$ are
also determined in terms of $S_{scat}$ only$.$ The origin of the second
requirement is the idea of an analytic change of ordering can be expressed in
terms of algebraic properties of emulats acting on particle states. The
underlying philosophy is similar to that in \cite{BKFZ}, one starts from an
analytic picture about analytic changes of $\theta$-orders in formfactors and
converts this with the help of emulation into an algebraic structure whose
validity can be directly checked. The crossing property results from the KMS
property together with this algebraic structure.

From \cite{BBS} we know that the operator properties of emulats in the general
case are radically different from those for integrable theories. Whereas for
the latter the plane wave formulation can be directly justified as a result of
the translation invariance of the domains of the emulats (this is the
intrinsic definition of "integrable"), the non-integrable emulats only exists
as $Z^{\#}(f)$ for $suppf\subset W,~$and their action is only defined on the
dense set of W-supported multi-particle states. Hence the above relations have
to be smeared with W-localized wave functions; without smearing they should
only be understood as relations for bilinear forms as in (\ref{bil}). Their
expected bad domain properties should result from the summation over
infinitely many intermediate states in the definition of $S_{gs}^{(n,m)}.$

What remains to be done is to show that the formal objects defined by the
above formulas can be backed up by operators $A_{in}(f)_{\mathcal{A}(W)}$
(\ref{ZF}) with the claimed domain properties and last not least that these
operators are wedge dual in the sense%

\begin{align}
&  \left\langle \psi\left\vert \left[  JA_{in}(\hat{f})_{\mathcal{A}%
(W)}J,A_{in}(\hat{g})_{\mathcal{A}(W)}\right]  \right\vert \varphi
\right\rangle =0,~J=S_{scat}J_{in}\label{weloc}\\
&  \left[  \mathcal{A}(W^{\prime}),\mathcal{A}(W)\right]  =0,~\mathcal{A}%
(W^{\prime})=\mathcal{A}(W)^{\prime}\equiv J\mathcal{A}(W)J\label{dual}%
\end{align}
which is the wedge duality (\ref{dual}) expressed in terms of the emulats. For
integrable models all these checks are straightforward and can be found in
\cite{AOP}\cite{Lech}, whereas in the general case they are only reasonably
simple in case one of the states is the vacuum $\left\vert \varphi
\right\rangle =\left\vert 0\right\rangle ~$(in which case the resulting
relation can be reduced to wedge duality in the free field algebra)$.$ We hope
to be able to complete the proofs in a future publication.

With such a difficult task still ahead, it is helpful to recall the aim of
this new nonperturbative setting. One objective is to show at least the
\textit{existence of models} which in view of being non-integrable can only be
\textit{approximated} in a controlled way. In the present setting one would
like to see if (in analogy to the integrable case) a crossing symmetric
unitary Poincar\'{e} invariant S-matrix determines a unique QFT. At least
under the two assumptions the answer is unique as it was already known before
for integrable models and conjectured for $S=1.$ The existence problem is not
answered by referring to the existence of wedge generators but it requires
another difficult step namely to show the nontriviality of double cone
intersection \cite{Lech}. Even in the integrable case this second step was
anything but simple \cite{Lech}.

There could be a chance that a kind of on-shell perturbation theory may lead
to a convergent perturbative construction of emulats together with the
S-matrix. One would e.g. start with a lowest order S-matrix in terms of the
mass-shell restriction of the interaction polynomial and use it in
(\ref{weloc}) to start an iteration whose first step is a lowest order emulat.
From these data one aims at the next order $S_{scat}$ and so on. Their may be
other ways to start an induction with low order emulates. The difference to
the Epstein-Glaser iteration for pointlike localized fields is that there are
no singular (operator-valued distributions) on-shell operators so that if the
singular structure (requiring renormalization) would have been the reason for
the divergence of the perturbative series, the on-shell situation in the
present case may be better. On shell perturbation theory in terms of the
S-matrix has been attempted at the time of the S-matrix bootstrap; from the
present setting it is clear that this remained without success since only
within a formfactor program one has sufficient structure (the matrix elements
if the $S_{scat}$-operator represent the formfactor of the identity operator)
to start an iteration.

The action of emulates on multiparticle states leads to a rather profound
insight into the validity of the crossing identity. It is clear that crossing
in the standard form%

\begin{align}
&  \left\langle 0\left\vert B\right\vert \theta_{1},.\theta_{k},\theta
_{k+1}.,\theta_{n}\right\rangle _{in}=~_{out}\left\langle \bar{\theta}%
_{k+1},.,\bar{\theta}_{n}\left\vert U(\Lambda_{W_{(0.1)}}(\pi i))B\right\vert
\theta_{1},.,\theta_{k}\right\rangle _{in}\label{cro}\\
&  B\in\mathcal{A(O}),~\mathcal{O}\subseteq W_{(0,1)},~\bar{\theta
}=antiparticle~of\text{ }\theta,~~\theta_{1}>..>\theta_{n}\nonumber
\end{align}
can only be valid if the particle states are smeared with nonoverlapping wave
functions $\hat{f}$\ so that the n-particle state can be expressed in terms of
a wave function-smeared naturally ordered product of $n$ $Z^{\ast}s$ applied
to the vacuum. In case of overlapping wave functions (the wedge localized wave
functions always overlap) we must use the above formulas for the action of
emulats in order to disentangle the action of the emulat on the left hand side
(\ref{id}) into particle states; in this process particle states of arbitrary
high particle number may enter. In fact without this complication the
emulation would be trivial and the KMS relation (\ref{K}) would be
indistinguishable from that of a free field. Even in the \textit{integrable
case} the contributions from the interaction modify the LSZ reduction formulas
e.g. \cite{BKFZ}%
\begin{align}
&  \left\langle \bar{\theta}\left\vert B\right\vert \theta_{1},..\theta
_{n}\right\rangle _{in}=\left\langle 0\left\vert B\right\vert \theta
_{1},..\theta_{n},\theta-i\pi\right\rangle _{in}+\\
&  +\sum_{j=1}^{n}\delta(\bar{\theta}-\theta_{j})\left\langle 0\left\vert
B\right\vert \theta_{1},.\hat{\theta}_{j}..\theta_{n}\right\rangle _{in}%
S_{gs}^{(j)}(\theta;\theta_{1},..\theta_{j-1})\nonumber
\end{align}
hence only the j=1 ($S_{gs}^{(1)}=1$) contact term is equal to what one
obtains from the naively derived LSZ reduction formula (which ignores the
modifications of overlapping situations). Only if the outgoing rapidities are
"smaller" (smaller $\theta s$) than those of incoming $\theta\not s  $ the
crossing relation takes the standard form; this is in particular fulfilled if
one starts from a vacuum formfactor (\ref{cro}). For higher dimensional QFT
the rapidity ordering has to be replaced by the velocity ordering with respect
to the wedge region $W~$\cite{BBS}$.$

Threshold singularities which lead to a breakdown of the Haag-Ruelle
scattering theory and to complicated changes of the LSZ reduction formula are
absent when neither the incoming and outgoing wave functions overlap among
themselves nor the incoming overlap with the outgoing situation. Hence the
crossing relation which connects the vacuum to n-particle formfactor with the
$k$ to $n-k$ particle formfactor holds since the ordering requirement can
always be fulfilled, but if one starts from a general formfactor there will be
complicated threshold modifications if the outgoing configutation is not
"smaller" than the incoming.

Hence the important message is that, although the crossing identity is related
to the KMS identity of wedge localization, one needs to know the action of the
emulats on multiparticle states in order to understand the details of that
relation. As in the old days of the (abandoned) bootstrap formalism the
\textit{crossing property is part of a constructive setting}. This role goes
much beyond that at the time of the dispersion relations. It is part of a new
design in which the second quantization functor, which leads from Wigner's
representation theoretical approach to interaction-free QFT, is replaced by
emulation of free fields (particles) within the wedge-localized interacting
operator algebra. In contrast to Lagrangian quantization, which works on a
parallelism to classical field theories, the emulation approach is totally intrinsic.

Being the formfactor of the identity operator, the S-matrix plays a special
role in this construction.
\begin{equation}
\left\langle \vartheta_{1},..\vartheta_{m}\left\vert S\right\vert \theta
_{1},..\theta_{n}\right\rangle =~_{out}\left\langle \vartheta_{1}%
,..\vartheta_{m}\left\vert \mathbf{1}\right\vert \theta_{1},..\theta
_{n}\right\rangle _{in}%
\end{equation}
As a result of the energy-momentum conservation it is not possible for the
important case of $2\rightarrow2$ particle scattering to cross a single
particle, one rather has to cross a pair, one from the incoming and one from
the outgoing configuration. The crossing property for pair crossing for the
elastic S-matrix is the only case for which the necessary analytic properties
were derived already at the time of the dispersion relations \cite{BEG}. The
authors applied the complicated theory of analytic functions of several
variables; this method did not reveal much about the conceptual environment of
crossing and according to my best knowledge it was not extended to other cases.

In the integrable case the analytic change leads to a representation of the
permutation group i.e. the analytic change does not depend on the path on
which it was carried out \cite{Lech}. This is radically different in the
general case; the above change involving the grazing shot S-matrix only holds
for the direct path (directly pulling an $Z^{\#}$ operator through a cluster
without zigzagging). This suggests analogies with other cases in QFT for which
operator changes correspond to analytic changes of positions in correlation
functions. In a QFT of Wightman correlation functions with standard spacelike
commutation relations the analytically continued correlation functions are
uni-valued in the Bargman-Hall-Wightman domain. On the other hand in a d=1+2
QFT with plektonic (braid group) commutation relation \cite{B-M} the
uni-valuedness in the BHW domain breaks down and the analytically continued
correlation depends on which analytic \textit{path} the coordinates have been
changed. Could there be a similar relation between generating fields of wedge
algebras constructed by emulation and their analytic changes in formfactors?
The algebraization in terms of emulats suggest that this is the case

The usual form of the crossing relates different formfactors while maintaining
the sum of bra +ket particles. It only holds if the wave functions of the ket
particles do not overlap those in the bra state. Otherwise the usual form
breaks down as a result of threshold singularities. In the general case the
analytic change couples the cluster of the crossed particles to \textit{all
other particles~}subject to the superselection rules. This amounts to an
on-shell realization of a radical form of Murphy's law: \textit{all that can
be coupled} (not forbidden by superselection rules) \textit{will be coupled}.
Integrable theories are precisely those which are protected against Murphy's law.

As in classical and quantum mechanics, an explicit construction of
nonintegrable QFT models is impossible. The aim of viewing emulation as a
generalization of Wigner's intrinsic representation theoretical approach to
the realm of interactions is to bring QFT to its conceptual closure by solving
its two remaining fundamental problems: proving existence of models (in
particular of physically interesting models) and finding mathematically
controllable approximation methods which could replace the standard diverging
perturbation series for correlation functions of renormalized fields.

The finiteness of on-shell objects was the main reason behind the various
S-matrix projects which arose in the late 50s as an attempt to extend the
successful derivation and experimental confirmation of dispersion relations
into a full S-matrix theory project. The present modular localization setting
shows why such attempts had no chance to succeed. An S-matrix construction
without its natural conceptual embedding as a relative modular invariant into
an approach based on modular theory of wedge localization does not present
enough structure to start computations; unitarity, Poincar\'{e} invariance and
the crossing property of an S-matrix are too general for providing a
computational basis. On the other hand ideas to enrich the S-matrix setting
with additional assumptions (as e.g. Mandelstam's postulated spectral
representation for scattering amplitudes) had a chance, at least as long as
they did not lead to outright contradictions with properties following from
the principles of local quantum physics.

As shown in the previous section, later S-matrix projects, as the dual model
and its string theory extension really did lead to such contradictions. The
present approach shows in a clear form that \textit{the crossing on which
Veneziano \cite{Ven} constructed his dual model has nothing to do with the
crossing in the sense of particle physics}. Rather what was called crossing in
the dual model referred to a kind of "field crossing" in Mellin transforms of
converging global operator expansions in conformal 4-point-functions
\cite{Mack} in which meromorphic functions arise, which have their first order
poles on a "dimensional trajectory" defined by scale dimensions of conformal
(composite) fields. It also makes no physical sense to subject Mellin
transforms to a unitarization process (interpreting the dual model as lowest
order of a new S-matrix construction), since the Hilbert space properties of
Mellin transforms and those from (approximations) of unitary S-matrices are
totally different.

Modular localization is also at odds with the idea of embedding a lower
dimensional QFT into a higher dimensional one. This is only possible in QM,
which has no intrinsic localization but only the one related to Born's
probability interpretation: a linear chain of oscillators does not "feel" the
space into which it is (it is up to the acting physicistn where to place it) embedded.

Proposals which are the result of a lack of foundational knowledge about local
quantum physics blended with an uncritical (if not to say messianic) use of
metaphoric analogies can carry a highly speculative science as particle
physics into a dead allay, unless a corrective critique rescues the situation
in time. I claim that this (without the rescue) is what exactly happened with
string theory and explains why after 5 decades theoreticians stand almost
empty-handed in front of LHC; those few individuals who tried to rescue the
situation were not listened to (had no status in a trend-driven world of
science). As in case of most sociological/political derailments, the wrong
path has to play out to the end, before something different can be started.

The present results may be summarized as follows. For integrable models the
bijective correspondence of wedge localized interacting operators leads to a
solution of the existence problem, so that the previously calculated
formfactors are really those of existing QFTs. In that case a description of
these operators as deformed free fields is very helpful \cite{Le}. Their
proximity to free fields reflects itself in the fact that in d=1+1 (and only
there) purely elastic S-matrix cannot be distinguished from an
interaction-free situation by cluster factorization. For non-integrable models
one only has a consistent constructive idea, the check of its correctness as
well as its usefulness requires more investigations.

In spite of the critical remarks about pure S-matrix constructions,
\textit{Mandelstam's step to place the S-matrix into a computational approach
was very important}, even if his later closing of ranks with dual model ideas
ended in a blind allay. Taking the original idea of using the S-matrix already
at the start of computations and combining it with Haag's idea of local
quantum physics (which led to modular localization), one obtains a powerful
new setting for a rigorously controlled particle theory project.

\subsection{Resum\'{e} and concluding remarks}

The success of the dispersion relation project led to the first
nonperturbative S-matrix based setting of particle physics. In form of the
S-matrix bootstrap it started from correct "axioms", its weakness was the lack
of an operational way of their implementation. Concepts as "maximal
analyticity" were conceptually as well as mathematically too vague and the
S-matrix bootstrap setting remained without concrete results.

The more constructive aftermath of the dispersion relation era started with
Mandelstam's conjectured two-variable representation, and moved via the dual
model to the canonically quantized Nambu-Goto Lagrangians and its d=1+9
dimensional supersymmetric infinite component QFT called "superstring". It was
more concrete and led to many detailed calculations but, as was shown in the
present work, \textit{it erred on the nature of the crossing property in
particle physics} and more generally got confused on the subtle issue of
quantum meaning of localization. Although QFT was born in the aftermath of the
Einstein-Jordan conundrum of fluctuation in a subvolume, whose solution
required the deep (and at that time not available) notion of modular
localization for its complete solution \cite{Ei-Jo} \cite{hol}, its age-old
incomplete understanding did not impede progress in perturbative aspects QFT.
The obvious reason is that renormalized perturbation theory is a "self-seller"
since, even in its most conceptual-mathematical form in terms of the
Epstein-Glaser approach, it can be implemented without knowing the deeper
intrinsic \textit{modular localization} aspects of QFT.

The return to a problem. which was for a very long time (since the very
discovery of QFT by Jordan) looming in the background outside of the
conceptual range, can be expected to lead to many other new perspectives in
particle theory, which could even bring about a different view about
quantization and perturbation theory. The motto of a paper written at the turn
of the century by Borchers \cite{Bo} with the title: \textit{On
revolutionizing quantum field theory with Tomita's modular theory}$~$is now
slowly becoming reality.

Even such a somewhat stagnating subject as gauge theory may get into gears
again. On the foundational level gauge theory (and any other QFT involving
massless higher spin potentials) from the point of view of Wigner
representation theory presents a \textit{deep clash of the Hilbert space
structure and modular localization}. The (m=0, s=1) Wigner representation
theory only contains (upon covariantization) generating pointlike $F_{\mu\nu
}(x)$ wave functions but there are no pointlike $A_{\mu}(x)$ which (as wave
function-valued distributions) generate the Wigner wave function space. But as
everybody was forced to accept, one cannot formulate interactions without
potentials i.e. only in terms of pointlike couplings of quantum matter with
field strengths. The quantization approach follows the Lagrangian quantization
\textit{parallelism to classical theory}, except that this clash has no analog
in the classical theory (i.e. $A_{\mu}^{class}(x)$ exists as any other
classical field). This leaves two possibilities in order to arrive at a QFT:
either quantize and loose the Hilbert space (calculate instead in an
indefinite space and return to a Hilbert space with the help of e.g. the BRST
formalism), or use covariant semiinfinite string-localized vector potentials
$A_{\mu}(x\not )  $ in Wigner space and try to introduce those objects into
interaction densities by extending the locality-based Epstein-Glaser setting.

Whereas the first approach does not contain any physical (off-shell) charged
matter fields, and one is forced to turn to momentum space recipes for
photon-inclusive cross sections for scattering of charged particles thus
sidelining a spacetime (off shell) understanding \ (as that involved in the
LSZ reduction formula) of charge-carrying objects in favour of a prescription
for momentum space photon-inclusive cross sections, the string-localized
approach in a Hilbert space opens other possibilities. One now has the chance
to understand the nonlocal quantum aspects of charged matter, which permits at
best a string-localized generating charged matter field, in terms of the
lowest order interaction.

This explains all those properties of charge carriers whose existence has been
previously inferred on the basis of structural arguments based on the quantum
Gauss law \cite{Haag} (which are in fact known to be in contradiction with the
formal aspects of the perturbative ghost formalism \cite{charge}%
\cite{unexplored}). The standard gauge approach only admits a charge neutral
BRST-invariant Hilbert space and one looses the chance to obtain a deeper
understanding of spacetime properties of charge carriers; even some charge
neutral objects (e.g. as two localized opposite charges with a gauge bridge
between them) are ,not part of the BRST formalism. The mechanism which
produces string-localized charge-carrying generating fields remains obscure
since the interaction density just looks like that of any other model which
couples pointlike fields and leads to pointlike interacting fields; the reason
why coupling of zero mass $s<1~$fields do not lead to infrared problems
whereas couplings of $s>1$ do remains unexplained.

The formalism based on string-localized potentials explains the
string-localization of charge carrying objects required by the quantum Gauss
law as a (perturbative) transfer of string localization from potentials to the
matter fields a process in which the potential gets away with still keeping
its associated pointlike field strength, the string localization of the charge
carrier cannot be undone by any linear operation.

The reformulation of gauge theory in terms of string-localization also leads
to a better understanding of the Schwinger-Higgs screening mechanism \cite{Sw}
which was known to experts in the 70s but this nice resentation of the Higgs
phenomenon was lost in the maelstrom of time, leaving behind the today's form
of the Higgs model with its sometimes misunderstood idea of a symmetry
breaking (which symmetry? Gauge is not a physical symmetry). We refer for the
description of the original Higgs model as arising from Schwinger-Higgs
charged screening of scalar QED as well as some speculative remarks about a
possibility to have a Higgs free massive YM model\footnote{Even if the
existence of the Higgs particle (i.e. the particle associated with the real
scalar field which survives after the charge of the complex scalar QED has
been screened) will be observed, it is always better to have theoretical
alternatives since less desperation enhances the credibility of results.} in
the string-localized setting to the existing literature \cite{unexplored}. The
string-localized version is the only formulation in which massive potentials
pass smoothly to their massless counterparts.

Another lost concept whose recollection could have changed the direction of
the dispute about the mathematical AdS-CFT correspondence (and prevented a
tsunami of thousands of publications on such a narrow subject without any
tangible result) is the Haag-Swieca-Buchholz result \cite{Haag} about phase
space degrees of freedom in QFT (mentioned in section 4). It plays an
important role once one goes beyond Lagrangian quantization, but still wants
to keep as much of its physical properties in a more intrinsic "axiomatic"
setting. With its deep (and not completely understood) relation to the causal
shadow property and the existence of global thermodynamic equilibrium states
\cite{Haag}, it is one of the pillars of Local Quantum Physics. Together with
modular localization these results are expected to become increasingly
important in more intrinsic settings of QFT beyond perturbation theory.

The progress on such issues led to a foundational understanding of integrable
models \cite{Lech}\cite{integrable} \cite{cross} and exposes for the first
time the limitations of the parallelism to classical physics known as
(Lagrangian, canonical) quantization which extend in both directions: for most
integrable models no Lagrangian is known (and not needed for their
construction) and there are families of mathematically well-defined classical
Lagrangians which have no counterpart in QFT\footnote{Any Lagrangian in a
$d>2$ with fields whose index space is noncompact does not have a quantum
counterpart unless the indices are tensor/spinor indices associated with the
Minkowski spacetime on which the field lives. Compact index spaces are allowed
and represent inner symmetries \cite{integrable}.}. The critique of string
theory in section 4 is closely related to such insights.

At this point a disquieting question comes to one's mind: have physicists
working in present day particle theory become less capable or has the subject
reached a dead end? It is my firm conviction that neither is true. What
happened is that the great progress in the first decades after worldwar II was
achieved with a relative simple theoretical investments which were sufficient
to discover important results, but were not quite adequate for their
foundational safeguarding within QFT, which would have been necessary to
obtain a strong conceptual mathematical platform for further innovative
explorations. As Feynman once expressed, innovative conquests often require to
bounce into the "blue yonder" but since many of such attempts end in failure
one needs a strong base to which one may safely return. The situation gets
completely out of hand if there is no such secure base and even reputable and
charismatic leaders of the meanwhile globalized scientific communities loose
their critical brakes. The result is a conceptual trap from which there is no
escape; even mathematical consistency may in the absence of a physical
conceptual guide lead to bizarre "results" \cite{Malda}. In this way the
vernacular "many people cannot err" is turned into its opposite. \ 

I have criticized the various pure S-matrix projects; but I tried to be
careful not to permit the inglorious end of the dual model/string theory to
invalidate the importance of Mandelstam's S-matrix based project which started
in the aftermath of the successful project of dispersion relations and the LSZ
scattering theory. By emphasizing the importance of the crossing property and
proposing a spectral representation for its exploration, he initiated a fresh
start which went beyond older failed attempts.

This was not always my belief, in fact as an adherent of LQP (local quantum
physics) I considered the S-matrix as an object which belongs to the roof and
not to the foundation of a QFT construction. There is some irony that, under
the influence of the bootstrap-formfactor program for factorizing models, I
had to change my view and take note that in the form of a \textit{relative
modular invariant of wedge localization} the S-matrix becomes part of the
constructive foundations. This point of view has meanwhile become the credo of
a small but growing community \cite{LQP} of highly dedicated young
researchers. In fact, I we argued above, it is this completely new view about
the foundational aspects of localization in QFT which sheds new light on LHC
relevant problems (as the Higgs issue) and renews the interest in old problems
which had already been forgotten in the maelstrom of (preelectronic) time.

\begin{acknowledgement}
am indebted to Herch Moyses Nussenzveig for valuable advice concerning
historical aspects.
\end{acknowledgement}

\end{document}